\documentstyle[aps,epsf]{revtex}

\begin{document}
\title{Hopping Conductivity of a Nearly-1d Fractal: a Model for Conducting Polymers}
\author{A.~ N.~ Samukhin$^{1,2}$, V.~N.~Prigodin$^{1,3}$, L.~Jastrab\'\i k$^2$, and
A.~J.~Epstein$^{3}$}
\address{$^1$A. F. Ioffe Physico-Technical Institute, 194021 St. Petersburg, Russia\\
$^2$Institute of Physics AS CR, Na Slovance 2, 180 40 Prague 8, Czech \\
Republic\\
$^3$Physics Department, The Ohio State University, Columbus, OH 43210-1106}
\maketitle

\begin{abstract}
We suggest treating a conducting network of oriented polymer chains as an
anisotropic fractal whose dimensionality $D=1+\epsilon $ is close to one.
Percolation on such a fractal is studied within the real space
renormalization group of Migdal and Kadanoff. We find that the threshold
value and all the critical exponents are strongly nonanalytic functions of $%
\epsilon $ as $\epsilon \rightarrow 0$, e.g., the critical exponent of
conductivity is $\epsilon^{-2}\exp (-1-1/\epsilon )$. The distribution
function for conductivity of finite samples at the percolation threshold is
established. It is shown that the central body of the distribution is given
by a universal scaling function and only the low-conductivity tail of
distribution remains $\epsilon $-dependent. Variable range hopping
conductivity in the polymer network is studied: both DC conductivity and AC
conductivity in the multiple hopping regime are found to obey a quasi-1d
Mott law. The present results are consistent with electrical properties of
poorly conducting polymers.
\end{abstract}

\draft

\section{Introduction.}

\label{intro}

Charge transport in structures with fractional dimensionality has attracted
a high degree of attention due to both its fundamental and applied interest 
\cite{mbb82,fp90}. Conductivity of a random fractal of resistors was studied
in Ref. \cite{k77} as an example of critical phenomena. A large body of
literature has been devoted to a study of the unusual dynamics of electrons
in regular and random fractals \cite{bh91}. The discussion of the questions
and appropriate references may be found in recent reviews (see, e.g. \cite
{hb87,nyo94}).

In the present paper we consider problems of percolation and hopping
transport on {\em nearly one-dimensional} strongly anisotropic fractals \cite
{spj97} with dimensionality $D=1+\epsilon $. These fractals are expected to
exhibit unique properties because their dimensionality at $\epsilon \ll 1$
is close to the low marginal one for the percolation transition. In contrast
to isotropic fractals a small parameter $\epsilon $ enables us to obtain the
exact solution. Moreover it becomes possible in the case of
nearly-one-dimensional fractals to establish not only average
characteristics but their entire distribution functions.

Another motivation for a study of quasi-one-dimensional fractals is recent
experimental data on conducting polymers such as doped polyacetylene,
polypyrrole and polyaniline \cite{ts92,icsm96,wjrme90,joea94,ke97}. In
general, this class of polymers has a great variety of transport properties.
In the undoped state these polymers are semiconductors with an energy gap of
Peierls-Mott origin \cite{Heeger88}. With doping the energy gap is
suppressed quickly and for the highly doped case there is finite density of
states at the Fermi level. The room temperature conductivity ($\sigma_{RT}$)
of heavily doped sample may attain metallic values and temperature and
frequency dependencies of conductivity may be close to metallic. The nature
of the metallic phase in these samples is presently a subject of intensive
study. One point is that the metallic state is dependent upon by strong
interchain coupling \cite{wjrme90,joea94,ke97,nps89}.

In doped polymers with moderate $\sigma_{RT}$ (of the order of several
hundreds $S/cm$) the conductivity, as a rule, decreases with decreasing
temperature \cite{ts92,wjrme90,joea94,ke97}. Because this decay follows a
power law in a large temperature interval, presumably, these materials are
near the metal-insulator transition which happens at the critical interchain
coupling. Poorly conducting doped samples with $\sigma_{RT}$ of the order or
less 1 $S/cm$ have behavior that can be classified as ``dielectric'' \cite
{wjrme90}: it is similar to that observed in amorphous semiconductors. For
such materials, DC conductivity is strongly dependent on temperature and,
generally, follows: $\sigma _{DC}\propto \exp -(T_0/T)^{1/2}$. It is noted
that caution must be made to review a large temperature range in comparison
of experimental conductivity with the model dependencies \cite{ke97}.

For a variable range hopping (VRH) mechanism of transport the temperature
dependence of conductivity was initially derived \cite{md79,se84} to be : $%
\sigma _{DC}\propto \exp -(T_0/T)^{1/(d+1)}$, where $d$ is system's
dimensionality. For $d=1$ this formula reproduces the observable dependence $%
\sigma_{DC}\propto \exp -(T_0/T)^{1/2}$. However this approach is not
correct, since the 1d VRH \cite{kk73} yields the Arrhenius law, $%
\sigma_{DC}\propto \exp -T_0/T$ with $T_0$ set by the highest barrier that
occurs in the chain. The 1d VRH was modified for a quasi-1d system \cite
{nps89,shante77} to include weak hops between the nearest-neighbor, thereby
avoiding the highest barriers. This approximation results in a quasi-1d VRH
law, $\sigma _{DC}\propto \exp -(T_0/T)^{1/2}$.

Experimental measurements of microwave conductivity and dielectric constant
in poorly conducting doped samples \cite{wjrme90,joea94} revealed that both
are strongly dependent upon temperature too, most probably according to the
same quasi-1d Mott's law, i.e. $\exp -(T_0/T)^{1/2}$. The usual theory of
hopping transport predicts, however, only a very weak power temperature
dependence for the frequency-dependent conductivity and the dielectric
constant in two-- and three--dimensional systems~\cite{bb85}.

In the present work we exploit the specific structure of the polymer network
to understand these peculiar features of conducting polymers. In stretched
polyacetylene this network is formed by coupled polymer chains oriented
along some direction. Electron micrographs shows that in these substances
polymeric chains are organized into {\em fibrils } \cite{ts92}, which may be
distinctly seen to be subdivided into smaller ones \cite{ar87}. In a
non-fibrillar form of conducting polymers, like polyaniline, X--rays data
reveal the existence of highly ordered ``crystalline regions'' with metallic
properties \cite{wjrme90,joea94,ke97}. Therefore the whole network of
stretched polyaniline may be thought of as constructed from long
one--dimensional polymer chains randomly coupled by metallic islands of
various sizes. The volume fraction of metallic islands can be small.

We assume here that polymer structure represents a {\em nearly
one-dimensional fractal}. That means a specific kind of polymer chain
organization, defined in the following way: Choose a three-dimensional cube
with the edge $L$. Chains, which are coupled within this cube, form a set of
bundles disconnected from each other. If for large enough $L$ the
cross--section of the maximum bundle is proportional to $L^\epsilon $, where 
$0\leq \epsilon \leq 2 $, then we shall call the system $d^{*}=1+\epsilon $
--dimensional. Obviously $\epsilon =0$ for purely one--dimensional systems
(sets of uncoupled chains). Note, that if one assume chains to be connected
either with a low concentration of uncorrelated interchain links, or with
weak links (their resistivities being high compared to intrachain ones in
our example), then we are dealing with a {\em \ quasi--one--dimensional}
system~ \cite{nps89}, which is three--dimensional according to our
definition.

The problem of electron localization in similar fractals was studied in
Refs. \cite{Shapiro82,Cohen88}. It was found that even in the presence of a
weak disorder all the electronic states remain localized as long as $%
\epsilon \le 1$. Therefore, a mechanism of charge transport in the fractal
with $\epsilon \ll 1$ is supposed to be variable range hopping (VRH). This
assumption is in agreement with the experimental observations for poorly
conducting highly doped polymers for which there is a finite density of
states at the Fermi level \cite{jpmme94}.

The usual method to treat VRH models is the effective medium approximation 
\cite{bb85}, which gives wrong results in the nearly--1d case. For example,
for the percolation model this method gives the threshold concentration of
broken bonds $c_t\approx \epsilon $, while $c_t\approx \exp \left(
-1/\epsilon \right) $, as we shall see later. The results for critical
exponents are also wrong in this approximation. To treat VRH in a nearly
one-dimensional fractal we choose the following approach. We will first
study the percolation problem in a nearly-1d fractal exactly. The VRH model
is reduced to the percolation problem by constructing the effective
percolation lattice \cite{bb85,bby79,zv80}.

In this way we have found that at low temperatures the VRH conductivity
obeys a quasi-1d Mott law: $\sigma_{DC}\propto \exp -(T_1/T)^{1/2}$ but the
characteristic temperature $T_1$ is greater than $T_0$ for 1d chain by a
factor $1/\epsilon$. Similar temperature dependence is obtained for AC
conductivity. These results can explain the observed temperature dependence
of conductivity and dielectric constant in poorly conducting polymers.
Additionally it was shown that there is the strong frequency dependence of
conductivity in the region of extremely low frequencies. These peculiarities
reflect the fact that in the random fractal with dimensionality close to one
the low frequency conductivity is entirely controlled by the weak charge
transfer between clusters. Each cluster is very dense and remains well
isolated.

There exist several different problems related to the percolation. First,
one can be interested in statistical properties of percolating media:
distribution of connected clusters, probability of two or more points to be
connected, etc. As it was shown by Fortuin and Kastelein \cite{fk72}, this
problem can be reduced to the $q$-component Potts model in the limit $%
q\rightarrow 1$. Thus, the powerful set of field theory methods may be
applied. This analogy, however, does not allow us to treat the conductivity
of a percolating cluster. The evaluation of the conductivity exponent $\mu $%
, which describes the DC conductivity $\sigma$ behavior near percolation
threshold: 
\begin{equation}
\sigma \propto \left( \frac{c_t-c}{c_t}\right) ^\mu \,,  \label{condexp}
\end{equation}
is much more complicated task than the ``field-theoretic'' ones, such as the
exponents of correlation length and of infinite cluster capacity, etc.

Thus our first aim is to study the critical behavior of conductivity near
the percolation threshold in a $d$--dimensional lattice, where $d$ is close
to lower critical dimensionality, i.e. $d=1+\epsilon $, $\epsilon \ll 1$.
The real space renormalization group of Migdal and Kadanoff (RGMK) \cite
{mg76,kdho75,kd76,tm96}, being exact at $d=1$, may be expected to be the
appropriate tool as $d$ tends to unity. This method was applied to the
percolation problem several years ago by Kirkpatrick \cite{k77}. He had
found critical exponents of correlation length and conductivity by using the
RG equations for conductivity distribution truncated up to the first moment.
Though he had not considered explicitly a case of nearly-1d system, this
method, if properly applied, gives the right dependence of conductivity
exponent on $\epsilon $ except for a pre-exponential factor.

We extend the RGMK method to consider the conductivities and resistivities
distribution functions in random media. This enable us to derive the
equation for the conductivity exponent $\mu$, which gives realistic values
not only in the nearly-1d case. Moreover, explicit expressions for the
distribution functions at the percolation threshold will be obtained. To the
best of our knowledge only numerical estimates of random conductivity
momenta were available until this work (see, e.g. \cite{bhl95}).

Thorough investigation of the percolation problem and its various
modifications is of interest both for its conceptual significance and due to
its numerous applications \cite{nyo94,imb92,sa94}. Beside its application to
random conducting media on which we shall concentrate here, the percolation
approach was used, e.g., to treat the rigidity transition in random networks 
\cite{mdl96}, and mechanical breakdown in solids \cite{zrsv96}. Another
application is the magnetic flux flow in type II superconductors \cite{wf96}%
. In the case of magnetic vortices pinned by disorder, their motion just
above the depinning threshold happens along a sparse (possibly, fractal)
network of persistent channels. Directed percolation model is often applied
now to describe a wide class of phenomena, in particular, self-organized
criticality \cite{gp95,mz96}.

The paper is organized as follows: In Chapter \ref{dimen} the notion of
fractional dimensionality is introduced for the oriented chain arrays and
illustrated by hierarchical structures. In Chapter \ref{rgmk} RGMK
transformations for conductivity of disordered media and for connectivity of
percolation system are derived, the latter one is studied in Chapter \ref
{connect}. The RG equation for the distribution function of conductivity at
the percolation threshold is solved in Chapter \ref{conduct}, and the
explicit form of distribution function is found in Chapter \ref{DF}. Using
scaling relations, results for the AC conductivity of the percolating
lattice near threshold are obtained in Chapter \ref{scale}, and they are
applied to describe the temperature and frequency dependence of conductivity
for the variable range hopping transport in Chapter \ref{vrh}. Results are
discussed in Chapter \ref{concl}. Three Appendixes contain technical details.

\section{Hierarchical coupling of oriented chains}

\label{dimen}

\begin{figure}[tbp]
\epsfbox{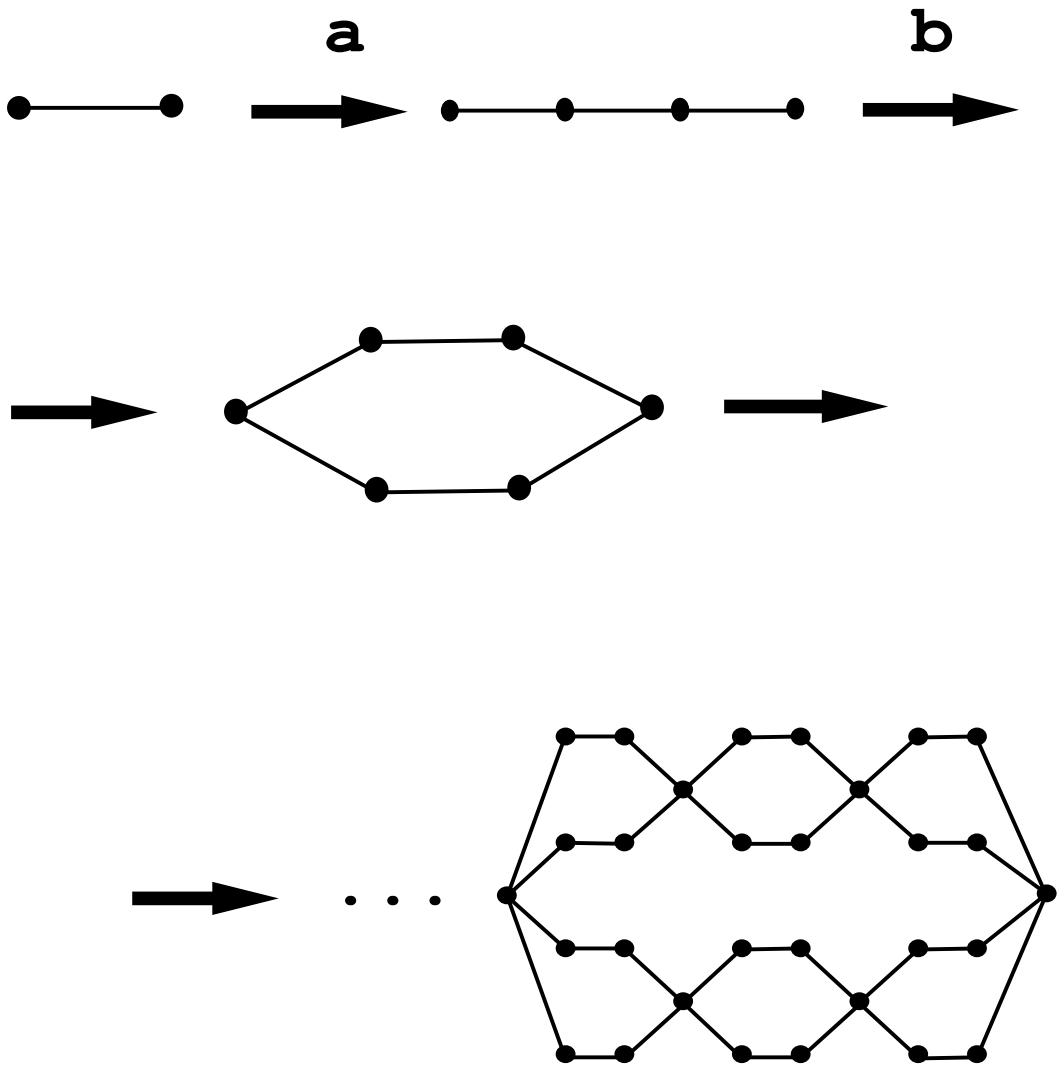}
\vspace{0.5cm}
\caption{Construction of hierarchical structure.}
\label{hs}
\end{figure}

The $(m,n)$ hierarchical structure (HS) is constructed through the infinite
repetition of two successive steps (see Fig. \ref{hs}): a) construction of $n
$-chains, and b) construction of $m$-bundles. After every step the resulting
construction may be treated as a new bond ($l$-th level bundles).

We use the following definition of dimensionality for a chain fractal
considering an array of one-dimensional chains, connected in some regular or
random fashion by transverse bonds of various lengths. In an $L$-size cube
chains form a set of bundles, connected inside this cube. Within each bundle
in the cube the chains are interconnected. There are no connection between
the bundles within the $L$-size cube. If the number of chains in the maximum
sized bundle scales as $L^\epsilon $ for large enough $L$, where $0\leq
\epsilon \leq 2$, then we have $D=1+\epsilon $ --dimensional network.
Obviously $\epsilon =0$ for a purely one--dimensional systems (sets of
disconnected chains). The characteristic feature of the fractals,
constructed from oriented 1d chains is their self-similarity: the system at
any scale looks like subdivided into bundles, which in turn are subdivided
into smaller ones, etc.

In particular, the dimensionality $D=1+\ln m/\ln n$ may be ascribed to the $%
(m,n) $ hierarchical structure, if in spatial dimension $d\ge D$ we replace
every site with $2m^l$ bonds attached to it ($l$ is the level of bundles
attached at each side of this site) with $m^l$ sites connected by transverse
bonds of infinite strength (see example in Fig. \ref{hsin2d}).

Our hypothesis here is that oriented polymer network structures are of this
type (with $D=1+\epsilon $ close to 1, $\epsilon \ll 1$), at least in some
wide enough interval of length scales, e.g. from the scale of fibrils
(hundreds of nm) down to molecular scales. Transmission electron micrographs
of fibrillar polyacetylene (see e.g. \cite{ar87}) appear to support this
hypothesis. Of course, the real structures are not regular ones, and the
requirement of self-similarity here is to be treated in statistical sense.
Nevertheless, we shall use the RGMK scheme, based on regular fractals (HS)
for their analysis. In the case of conducting polymers such as nonfibrillar
doped polyaniline and doped polypyrrole one may assume the polymer networks
apparent fractality in some scale range to be caused by a dilute
distribution of crystalline regions providing interchain links (fractality,
generated by randomness \cite{hba96}). Having in mind that the RGMK is exact
in one dimension, one may hope to obtain meaningful results when the
dimensionality is close to 1.

\section{Migdal and Kadanoff equations}

\label{rgmk}

The renormalization group of Migdal and Kadanoff may be formulated in a
quite simple phenomenological fashion. Suppose we have some random $D$
-dimensional medium with fluctuating local conductivity/resistivity. Let us
consider a $\lambda $-size cube within the medium. Its conductance is $%
\sigma (\lambda )\lambda ^{D-2}$, where the random conductivity $\sigma
\left( \lambda \right) $ is $\lambda $-dependent for strongly inhomogeneous
systems. The distribution function for conductivity and for resistivity $%
\rho(\lambda )=1/\sigma (\lambda )$ are defined in the Laplace
representation as: 
\begin{equation}
P(\eta ,\lambda )=\left\langle \exp \left( -\eta \sigma \left( \lambda
\right) \right) \right\rangle ;\quad Q(s,\lambda )=\left\langle \exp \left(
-s\rho\left( \lambda \right) \right) \right\rangle .  \label{def1}
\end{equation}

\begin{figure}[tbp]
\epsfbox{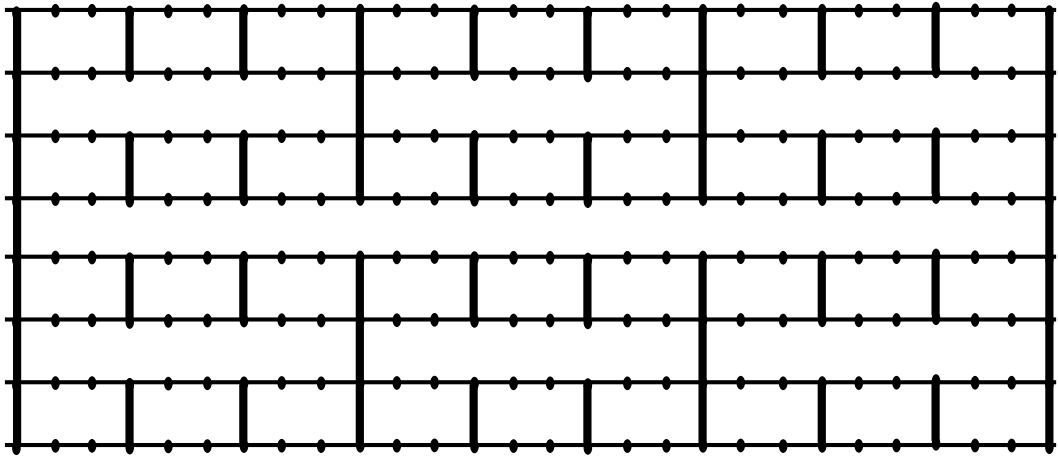}
\vspace{0.5cm}
\caption{Hierarchical structure of Fig. \ref{hs} depicted for a
two-dimensional case.}
\label{hsin2d}
\end{figure}

If we change the size of the cube, $\lambda \rightarrow \lambda ^{\prime
}=n\lambda $, we arrive at some new random variables $\sigma (\lambda
^{\prime })$, $\rho(\lambda ^{\prime })$ with distribution functions $%
P\left( \eta ,\lambda ^{\prime }\right) $, $Q\left( s,\lambda ^{\prime
}\right)$, respectively. The cube's enhancement may be treated as $n$-times
expansion in one (``longitudal'') spatial direction, and in $D-1$ other
(``transverse'') ones. If one intends to treat these transformations as
infinitesimal ones afterwards, the order of operations is not important.

The RGMK scheme is based upon two approximations: i) enhancing the size $n$
times in the longitudinal direction is treated as connection of $n$
resistors in series, and ii) in a similar way, the transverse cube's
enhancement is replaced by the parallel connection of $m=n^{D-1}$ elements.
Thus we have: 
\begin{eqnarray}
\widetilde{\rho}^{\left( n\right) }\left( \lambda \right) =\frac 1n
\sum_1^n\rho_l\left( \lambda \right)\,,\,\,\,\,\sigma \left( n\lambda
\right) =\frac 1m\sum_1^m\widetilde{\sigma }_l^{\left( n\right)} \left(
\lambda \right)\,.  \label{adlaw}
\end{eqnarray}
Here the tilde values refer to the rectangular element with the dimensions $%
n\lambda $ in the longitudinal direction and $\lambda $ in other ones. In
both steps the resistivities $\rho_l$ and conductivities $\widetilde{\sigma}%
_l^{\left(n\right)}=1/\widetilde{\rho}_l^{\left(n\right)}$ of the
constituent component are supposed to be independent random variables, and,
therefore, we have the distribution functions to be transformed in two steps
simply as: 
\begin{equation}
\tilde{Q}(s,\lambda )=\left\langle \exp \left( -s\tilde{\rho}^{\left(
n\right) }\left( \lambda \right) \right) \right\rangle =Q^n\left(
s/n,\lambda \right) \,,\;P\left( \eta ,n\lambda \right) =\tilde{P}^m\left(
\eta /m,\lambda \right),~~~~m=n^{D-1}.  \label{lgtr}
\end{equation}
This transformation is exact for the $(m,n)$ hierarchical structure.

Equation (\ref{lgtr}) should be supplemented with the relation between
conductivities and resistivities distribution functions (DF) in the Laplace
representation. It may be easily derived from the definitions (\ref{def1}),
using the following integral identity: 
\[
e^{-x/\alpha }=1-\sqrt{x}\int_0^\infty \frac{dy}{\sqrt{y}}J_1\left( 2\sqrt{%
xy }\right) e^{-\alpha y}\,, 
\]
where $J_1$ is the Bessel's function. As a result, the relation between $%
Q(s,\lambda)$ and $P(\eta,\lambda)$ takes the form of Hankel's
transformation: 
\begin{equation}
Q(s,\lambda )=1-\sqrt{s}\int_0^\infty \frac{d\eta }{\sqrt{\eta }} J_1\left( 2%
\sqrt{s\eta }\right) P\left( \eta ,\lambda \right) \,,  \label{hankel}
\end{equation}
the reverse relation is of the same form. This transformation has the
following properties to be used later: 
\begin{equation}
Q\left( 0,\lambda \right) =1-P\left( +\infty ,\lambda \right) \,,\quad
Q\left( +\infty ,\lambda \right) =1-P\left( 0,\lambda \right) \,,  \label{ha}
\end{equation}
\begin{equation}
s\frac{\partial Q(s,\lambda )}{\partial s}=\sqrt{s}\int_0^\infty \frac{
d\eta }{\sqrt{\eta }}J_1\left( 2\sqrt{s\eta }\right) \eta \frac{ \partial
P\left( \eta ,\lambda \right) }{\partial \eta }\,,  \label{hd1}
\end{equation}
\begin{equation}
s\frac{\partial ^2Q(s,\lambda )}{\partial s^2}=\sqrt{s}\int_0^\infty \frac{
d\eta }{\sqrt{\eta }}J_1\left( 2\sqrt{s\eta }\right) \eta P\left( \eta
,\lambda \right) \,\,.  \label{hd2}
\end{equation}

Thus we have a closed set of equations for arbitrary rescaling factors $n$
and $m$. It appears to be more convenient to deal with an infinitesimal
transformation by setting $n=1+\delta \lambda /\lambda $, $m=1+\epsilon
\delta \lambda /\lambda $. From Eq.(\ref{lgtr}) we have the variation of
distribution functions consisting of longitudinal and transverse parts: $%
\delta P=\delta _lP+\delta _tP$, $\delta Q=\delta _lQ+\delta _tQ$. 
\begin{eqnarray}
\delta _lQ(s,\lambda ) &=&\left[ -s\frac{\partial Q(s,\lambda )}{\partial s}
+Q(s,\lambda )\ln Q(s,\lambda )\right] \frac{\delta \lambda }\lambda \,, 
\nonumber \\
\delta _tP(\eta ,\lambda ) &=&\epsilon \left[ -\eta \frac{\partial P(\eta
,\lambda )}{\partial \eta }+P(\eta ,\lambda )\ln P(\eta ,\lambda )\right] 
\frac{\delta \lambda }\lambda .  \label{variz}
\end{eqnarray}
Using the relations (\ref{hankel},\ref{hd1}), one may rewrite the first of
these equations as: 
\begin{equation}
\delta _lP(\eta ,\lambda )=\left[ \eta \frac{\partial P(\eta ,\lambda ) }{%
\partial \eta }-\int_0^\infty \frac{ds}{\sqrt{s}}J_1\left( 2\sqrt{ s\eta }%
\right) Q(s,\lambda )\ln Q(s,\lambda )\right] \frac{\delta \lambda } \lambda
\,.  \label{varlg}
\end{equation}
Then, from $\delta P=\frac{\partial P}{\partial \lambda }\delta \lambda
=\delta _lP+\delta _tP$, we have the following equation: 
\begin{eqnarray}
&&\lambda \frac{\partial P\left( \eta ,\lambda \right) }{\partial \lambda }
=B\left( \left\{ P\right\} ,\eta \right) =  \nonumber \\
&&\left( 1-\epsilon \right) \eta \frac{\partial P(\eta ,\lambda )}{ \partial
\eta }+\epsilon P(\eta ,\lambda )\ln P(\eta ,\lambda )-\int_0^\infty \frac{ds%
}{\sqrt{s}}J_1\left( 2\sqrt{s\eta }\right) Q(s,\lambda )\ln Q(s,\lambda )\,.
\label{eveq}
\end{eqnarray}
This equation, combined with the Eq. (\ref{hankel}), determines the
evolution of the distribution function upon size scaling in a closed form.

This scheme also allows us to treat the percolation system, introducing the
probability that the $\lambda$-sized cube is disconnected (i.e., has zero
conductivity or infinite resistivity) $c(\lambda )$. Taking into account the
definitions of distribution functions (\ref{def1}), $c(\lambda)$ may be
written as: 
\begin{equation}
c(\lambda )=P\left( +0,\lambda \right) =1-Q\left( +\infty ,\lambda \right)
\,.  \label{bb}
\end{equation}
Putting in Eq. (\ref{eveq}) $\eta =+\infty $ and using the formula (\ref{ha}%
), we have: 
\begin{equation}
\frac{dc}{d\lambda }=\epsilon c\ln c-\left( 1-c\right) \ln \left( 1-c\right)
\,.  \label{bbev}
\end{equation}
The right hand side of this equation has three fixed points: two stable
ones, $c=0$ and $c=1$, corresponding to connected and disconnected systems,
respectively, and the unstable fixed point, $c=c_t$, $0<c_t<1$, 
\begin{equation}
\epsilon c_t\ln c_t=\left( 1-c_t\right) \ln \left( 1-c_t\right) \,,
\label{thrp}
\end{equation}
corresponding to the percolation threshold.

Now let us consider statistical properties of clusters for the percolation
problem, i.e., the distribution of clusters over sizes and site numbers,
existence and capacity of the infinite cluster, etc. We suppose every bond
of a HS, to be either broken, with probability $c$, or not, --- with
probability $1-c$. The statistics of percolating network is closely related
to the thermodynamics of the $q$-states Potts model \cite{fk72}. Namely, if
we consider the partition function of the latter: 
\begin{equation}
Z_q=\exp \left( -{\cal H}_q^{(0)}\right) ,  \label{pf}
\end{equation}
\begin{equation}
{\cal H}_q^{(0)}=K\sum_{\left\langle ij\right\rangle }\left( 1-\delta _{\eta
_i\eta _j}\right) ,  \label{hp0}
\end{equation}
where variables $\eta _i=0,1,\ldots ,q-1$, $\delta $ is the Kronecker $%
\delta $-symbol. $Z_q$ may be expressed in terms of the percolation model
as: 
\begin{equation}
Z_q=\left\langle q^\Gamma \right\rangle \,.  \label{pfp}
\end{equation}
Here $\Gamma $ is the total number of connected clusters in the percolation
model, and the average is over realizations with the broken bonds
concentration $c=\exp (-K)$.

To establish a further relationship, it is necessary to introduce external
fields into the Hamiltonian of the Potts model: 
\begin{equation}
{\cal H}_q^{(1)}={\cal H}_q^{(0)}+h_1\sum_i\left( 1-\delta _{\eta
_i,0}\right) ,  \label{hp1}
\end{equation}
which may be thought of as additional bonds with strength $h_1$ between
every site of a given lattice and some fictitious external one.

The statistical properties of the percolation model may be obtained from its
``free energy'' 
\begin{equation}
f\left( K,h_1\right) =-\frac 1N\left. \frac \partial {\partial q}\ln
Z_q\left( K,h_1\right) \right| _{q=1}=-\frac 1N\left\langle \Gamma
\right\rangle _{h_1}\,,  \label{fe}
\end{equation}
where $N$ is the total number of sites. For an example, the order parameter 
\begin{equation}
P_1\equiv 1-\left. \frac{\partial f}{\partial h_1}\right| _{h_1=0}\,,
\label{op1}
\end{equation}
characterizes the ``capacity'' of the infinite cluster, i.e., the
probability of a given site to belong to the infinite cluster. The second
derivative is the ``susceptibility'': 
\begin{equation}
\chi =-\left. \frac{\partial ^2f}{\partial h_1^2}\right| _{h_1=0}\,,
\label{suc}
\end{equation}
which gives the average number of sites in finite clusters \cite{nyo94}.

To obtain the RG equation for $f$ of a HS, one should sum over intermediate
sites in bundles, thus, performing a transition from the initial
Hamiltonian, containing variables of all sites, to the one, containing
variables of sites of the next level (see Fig. \ref{hsin2d}). This
transformation reproduces the structure of the initial Hamiltonian (\ref{hp1}%
) with an additional term, corresponding to an extra external field $h_2$: 
\begin{equation}
{\cal H}_q={\cal H}_q^{(1)}-h_2\sum_{\left\langle ij\right\rangle }\left[
\delta _{\eta _i,0}\delta _{\eta _j,0}-\delta _{\eta _i\eta _j}- \frac 12%
\left( \delta _{\eta _i,0}+\delta _{\eta _j,0}\right) +1\right] \,,
\label{hp}
\end{equation}
The last two terms in the sum are introduced for further convenience.

For the $(n,m)$ HS, due to its self-similarity, after the summation over
variables at intermediate sites we have the following equality: 
\begin{equation}
Z_q(K,h,N)=\exp \left[ \frac N{mn}f_q^{(0)}(K,h)\right] Z_q\left( K^{\prime
},h^{\prime },\frac N{mn}\right) \,,  \label{pftr}
\end{equation}
where $h\equiv (h_1,h_2)$. Expressions for the parameters of the new
Hamiltonian $K^{\prime }$ and $h^{\prime }$, and the function $f_q^{(0)}$
may be found using the transfer matrix formalism. Introducing the transfer
matrix for a bond: 
\begin{eqnarray}
T_{\eta _1\eta _2} &=&\exp \left[ -H(\eta _1,\eta _2)\right] \,,  \nonumber
\\
H\left( \eta _1,\eta _2\right) &=&K\left( 1-\delta _{\eta _1\eta _2}\right) +%
\frac{h_1}2\left( 2-\delta _{\eta _1,0}-\delta _{\eta _2,0}\right) - 
\nonumber  \label{trm} \\
&&h_2\left[ \delta _{\eta _1,0}\delta _{\eta _2,0}-\delta _{\eta _1\eta _2}-%
\frac 12\left( \delta _{\eta _1,0}+\delta _{\eta _2,0}\right) +1\right] \,,
\end{eqnarray}
it is easy to calculate the transfer matrix for the $(n,m)$ bundle: 
\begin{equation}
T_{\eta _1\eta _2}^{\prime }=\left[ \left( \hat T^n\right) _{\eta _1\eta
_2}\right] ^m=\exp \left[ f_q^{(0)}(K,h)-H^{\prime }(\eta _1,\eta _2)\right]
\,,  \label{ntrm}
\end{equation}
with $H^{\prime }$ having the same structure as $H$, but with new parameters 
$K^{\prime }$ and $h^{\prime }$. Then we come to the following equation for
the free energy of the Potts model: 
\[
f_q(K,h)\equiv -\frac 1N\ln Z_q=\frac 1{mn}f_q(K^{\prime },h^{\prime })- 
\frac 1{mn}f_q^{(0)}(K,h)\,, 
\]
and for the percolation model: 
\begin{equation}
f(c,h)=\frac 1{mn}f(c^{\prime },h^{\prime })-\frac 1{mn}u(c,h)\,,
\label{fetr}
\end{equation}
where the variable $c=\exp (-K)$ is introduced instead of $K$, and $%
u(c,h)=\partial f_q^{(0)}/\partial q|_{q=1}$.

Finally, after transition to infinitesimal transformations: 
\begin{eqnarray}
n &=&1+\frac{d\lambda }\lambda ,\quad m=1+\epsilon \frac{d\lambda }\lambda
\,,  \nonumber \\
c^{\prime } &=&c+v_c\frac{d\lambda }\lambda \,,\quad h_{1,2}^{\prime
}=h_{1,2}+v_{1,2}\frac{d\lambda }\lambda \,,\quad u=w\frac{d\lambda }\lambda
\,,  \label{izdef}
\end{eqnarray}
we arrive at the following equation: 
\begin{equation}
v_c\frac{\partial f}{\partial c}+v_1\frac{\partial f}{\partial h_1}+v_2\frac{
\partial f}{\partial h_2}-\left( 1+\epsilon \right) f=w\,,  \label{efe}
\end{equation}
where $v_c$, $v_{1,2}$ and $w$ are found in Appendix \ref{apa}, Eqs.(\ref{vc}%
-\ref{w}). Setting $h_{1,2}=0$ in Eq.(\ref{efe}), and taking into account
Eq.(\ref{w}), we have: 
\begin{equation}
\lambda \frac{df}{d\lambda }-(1+\epsilon )f=w_0=c+(1-c)\ln (1-c)\,,
\label{efe0}
\end{equation}
where the independent variable $\lambda $ is related to $c$ by Eq. (\ref
{bbev}).

Equation (\ref{efe}) should be supplemented by the boundary conditions.
Directly from the definition (\ref{fe}) we have: $f=-1$ at $c=0$ and $f=0$
at $c=1$. Both Equations (\ref{efe},\ref{efe0}) have essentially two
different solutions depending on $c>c_t$ or $c<c_t$, where $c_t$ is the
threshold concentration of broken bonds determined from Eq. (\ref{thrp}).

\section{Percolation exponents}

\label{connect}

In the present section the critical exponents of connectivity will be
obtained. Critical exponent $\nu $ of the correlation length $\xi $: 
\begin{equation}
\xi \propto |c-c_t|^{-\nu }\,,  \label{defnu}
\end{equation}
is determined by linearization of Eq. (\ref{bbev}) near $c=c_t$: 
\begin{equation}
\nu ^{-1}=\epsilon \left( 1+\ln c\right) +1+\ln (1-c).  \label{cle}
\end{equation}

Eqs.(\ref{bbev},\ref{efe0}) can be solved analytically for $\epsilon \ll 1$.
In this case we have from Eq.(\ref{thrp}): 
\begin{equation}
c_t=e^{-1/\epsilon }\,.  \label{thrn1d}
\end{equation}
After substituting Eq. (\ref{thrn1d}) into Eq. (\ref{cle}), the critical
exponent $\nu $ reads: 
\begin{equation}
\nu \approx \frac 1\epsilon.  \label{clen1d}
\end{equation}
The procedure for the solution of Eqs. (\ref{bbev},\ref{efe0}) is described
in Appendix \ref{apb}. The ``partition function'' $f(c)$ is found to be: 
\begin{equation}
f(c)=\left\{ 
\begin{array}{c}
-c,\quad c\gg c_t; \\ 
-\frac{c_t^2}{2\epsilon }\left( 2\ln \frac{c_t}c\right) ^{\frac 1\epsilon
+1}\gamma \left( -\frac 1\epsilon -1,2\ln \frac{c_t}c\right) -\epsilon ^{%
\frac 1\epsilon +1}\left( \ln \frac{c_t}c\right) ^{\frac 1\epsilon
+1}\,,\quad 1\gg c>c_t; \\ 
\frac{c_t^2}{2\epsilon }\left( 2\ln \frac{c_t}c\right) ^{\frac 1\epsilon
+1}\Gamma \left( -\frac 1\epsilon -1,2\ln \frac{c_t}c\right) \,,\quad c<c_t;
\end{array}
\right.  \label{fe0}
\end{equation}
where $\Gamma (a,x)$ is the incomplete $\Gamma $-function \cite{be}, $\gamma
(a,x)=\Gamma (a)-\Gamma (a,x)$. It is curious that the ``singular'' part of
the ``free energy'' to the left of the percolation threshold, $c<c_t$: 
\begin{equation}
f_s^{-}(c)=-\frac{\sqrt{\pi }}2c_t\frac{\left( 2\epsilon \right) ^{\frac 1 %
\epsilon +\frac 12}}{\sin \frac \pi \epsilon }\left( \ln \frac{c_t}c\right)
^{\frac 1\epsilon +1}  \label{fe0s}
\end{equation}
is strongly oscillating function of $\epsilon $ at $\epsilon \ll 1$. The
critical exponent for the ``specific heat'' $\alpha $, $f_s\propto
|c-c_t|^{2-\alpha }$ appears to be large negative (as usual in the
percolation model, see, e.g. \cite{nyo94}).

Differentiating Eq.(\ref{efe}) with respect to $h_{1,2}$ and setting $%
h_{1,2}=0$ (see also Eqs.(\ref{vc}-\ref{w})) we get equations for the
``order parameters'' $P_1$ (Eq.(\ref{op1})) and $P_2\equiv -\left. \partial
f/\partial h_2\right| _{h=0}$: 
\begin{eqnarray}
\lambda \frac{dP_1}{d\lambda } &=&\frac 1c\left[ 2-c+\frac 2c(1-c)\ln
(1-c)\right] cP_1-P_2\,,  \nonumber \\
\lambda \frac{dP_2}{d\lambda } &=&\left[ c+\ln (1-c)\right] P_1+\left[ 1-c+ 
\frac 1c(2-c)\ln (1-c)\right] P_2\,.  \label{ope}
\end{eqnarray}
Boundary conditions: $\left. m_1\right| _{c=0}=1$, $\left. m_2\right|
_{c=0}=0$, $\left. m_1\right| _{c=1}=\left. m_2\right| _{c=1}=0$ may be
obtained directly from the definition of order parameters. At $c>c_t$ we
have the trivial solution of Eqs. (\ref{ope}), $P_1=P_2=0$. The lowest
eigenvalue of the matrix in the right hand side of Eq.(\ref{ope}) at $c=c_t$%
, which also is $\beta /\nu $, may be easily found numerically for any
dimensionality. The results are presented in Table \ref{exponents} together
with ones obtained by other methods.

For the nearly-1d system one can find the explicit form of order parameter
dependence $P_1\left( c\right) $ and $P_1\left( \lambda \right) $. Because
the region of our interest is $c<c_t\ll 1$, Eqs. (\ref{ope}) can be
rewritten for small $c$ as: 
\begin{eqnarray}
\lambda \frac{dP_1}{d\lambda } &=&\frac{c^2}3P_1-\frac c3P_2\,,
\label{open1da} \\
\lambda \frac{dP_2}{d\lambda } &=&-\frac{c^2}2P_1+P_2\,.  \label{open1d}
\end{eqnarray}
One can see from Eq.(\ref{open1d}), that $P_2=O(c^2P_1)$, therefore one can
neglect the second term in Eq. (\ref{open1da}). As a result Eq. (\ref
{open1da}) directly is solved by the substitution $c=c_t\exp(-\lambda^%
\epsilon)$ (see Eq. (\ref{solc1})) to read: 
\begin{equation}
P_1=\exp \left[ -\frac{c_t^2}{3\epsilon }\int\limits_{\lambda ^\epsilon
}^\infty \frac{dx}xe^{-2x}\right] =\exp \left[ \frac{c_t^2}{3\epsilon }{\rm %
Ei}\left( -2\lambda ^\epsilon \right) \right] \,,  \label{op}
\end{equation}
where ${\rm Ei}$ is the integral exponent function. Taking into account the
asymptotics of ${\rm Ei}\left( -x\right) =\ln x-C+\dots $ at small $x$, and
the relation between $c$ and $\lambda $, near the percolation threshold we
obtain: 
\begin{equation}
P_1=\left( \frac{c_t-c}{c_t}\right) ^\beta \,,\quad \beta =\frac{c_t^2}{
3\epsilon }=\frac 1{3\epsilon }\exp \left( -\frac 2\epsilon \right) \,.
\label{beta}
\end{equation}
Thus the critical exponent $\beta $ of the infinite cluster capacity appears
to be very small and a strongly nonanalytic function of $\epsilon $.

Using the scaling relations\cite{nyo94} , the complete set of critical
exponents for the connectivity problems may be expressed through $\nu $ and $%
\beta $ already obtained.

\section{Conductivity exponent}

\label{conduct}

Let us consider in more details the properties of the evolution functional $%
B $ in Eq.(\ref{eveq}). If we assume, e.g., that the conductivity is 0, with
probability $c$, and equals some finite value (say, 1), with probability $%
1-c $, then its distribution function reads: $c+\left( 1-c\right) e^{-\eta }$%
. Choosing the distribution in Eq. (\ref{eveq}) close to this form: 
\begin{equation}
P\left( \eta \right) =c+\left( 1-c\right) e^{-\eta }+\psi \left( \eta
\right) \,,  \label{lin}
\end{equation}
where $\psi(\eta)$ is the small correction. Linearizing the evolution
operator $B$ with respect to $\psi $, we have: 
\begin{eqnarray}
\ &&B\left( \left\{ P\right\} ,\eta \right) =\epsilon \left( 1-c\right) \eta
e^{-\eta }+\epsilon \left[ c+\left( 1-c\right) e^{-\eta }\right] \ln \left[
c+\left( 1-c\right) e^{-\eta }\right] +  \nonumber \\
&&\left( 1-c\right) \ln \left( 1-c\right) \,\left( 1-e^{-\eta }\right) +\eta 
\frac{d^2\psi }{d\eta ^2}+\left( 1-\epsilon \right) \eta \frac{ d\psi }{%
d\eta }+\left\{ 1+\epsilon +\epsilon \ln \left[ c+\left( 1-c\right) e^{-\eta
}\right] \right\} \psi \,.  \label{linev}
\end{eqnarray}

Another approximation for $B$ is possible if we assume: 
\begin{equation}
P\left( \eta \right) =c+\left( 1-c\right) \exp \left[ -\phi (\eta )\right]
\,,  \label{defphi}
\end{equation}
with $\phi \left( 0\right) =0$, $\phi \left( \eta \right) \rightarrow \pm
\infty $ as $\eta \rightarrow \pm \infty $ rapidly enough (faster than $\pm 
\sqrt{\left| \eta \right| }$, as we shall see later).

\begin{figure}[tbp]
\epsfbox{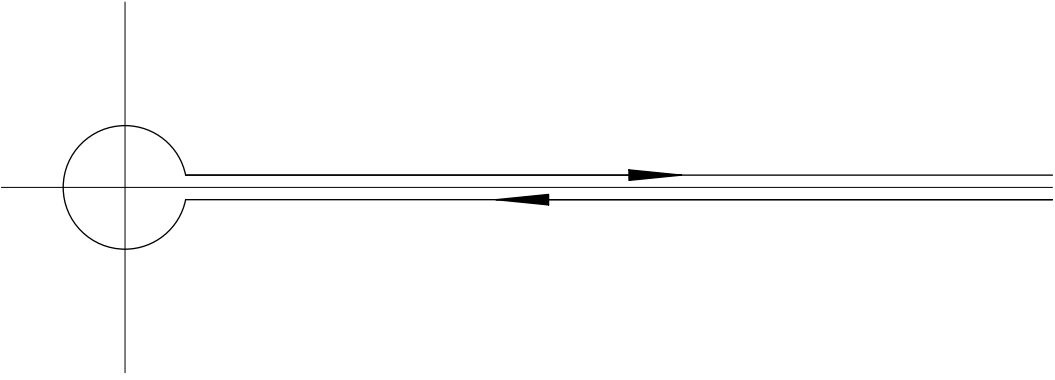}
\vspace{0.5cm}
\caption{Integration contour in Eq. (\ref{hank1}).}
\label{intc}
\end{figure}

The important point is also to assume analyticity of $P\left( \eta \right) $
and of $Q(s)$ within some stripe along the real axis. Using the relations: 
\[
J_1(z)=\frac{H_1^{(1)}(z)+H_1^{(2)}(z)}2\,,\quad H_1^{(1)}\left( ze^{i\pi
}\right) =-H_1^{(2)}(z)\,,\quad H_1^{(1)}(z)=-\frac{2i}{\pi z}\,\text{as }
z\rightarrow 0\,, 
\]
where $H_1^{(1,2)}$ are Hankel's function of first and second kind,
respectively, we may replace the integrals with the $J_1$-functions in Eqs. (%
\ref{hankel},\ref{eveq}) with the ones containing $H_1^{(1)}$, along the
contour $C$ shown on Fig. \ref{intc}. Thus we obtain: 
\begin{eqnarray}
Q(s) &=&\left( 1-c\right) \left[ 1-\sqrt{s}\int_0^\infty \frac{d\eta }{\sqrt{%
\eta }}J_1\left( 2\sqrt{s\eta }\right) \exp \left( -\phi \left( \eta \right)
\right) \right]  \label{hank1} \\
&=&-\left( 1-c\right) \frac{\sqrt{s}}2\int_C\frac{d\eta }{\sqrt{\eta }}
H_1^{(1)}\left( 2\sqrt{s\eta }\right) \exp \left( -\phi \left( \eta \right)
\right) ,  \label{hank1a}
\end{eqnarray}

where the contribution of the pole of $H_1^{(1)}\left( 2\sqrt{s\eta }
\right) /\sqrt{\eta }$ in Eq. (\ref{hank1a}) just reproduces the first term
in the square brackets of Eq. (\ref{hank1}). Assuming $\left| s\right| $ to
be large enough, one may replace $H_1^{(1)}$ by its asymptotic expression: 
\[
H_1^{(1)}\left( 2\sqrt{s\eta }\right) \simeq \pi ^{-1/2}\left( s\eta \right)
^{-1/4}\exp \left( -\frac{3i\pi }4+2i\sqrt{s\eta }\right) \,, 
\]
and treat the integral (\ref{hank1a}) by the saddle point method. As a
result, we get: 
\begin{equation}
Q\left( s\right) =\left( 1-c\right) \left[ \frac{\phi ^{\prime }(\eta _c)}{
\phi ^{\prime }(\eta _c)+2\eta _c\phi ^{\prime \prime }(\eta _c)} \right]
^{1/2}\exp \left[ -\phi (\eta _c)+2\eta _c\phi ^{\prime }(\eta _c)\right] \,,
\label{sp}
\end{equation}
where the saddle point should be determined from the equation: 
\begin{equation}
i\sqrt{\frac s{\eta _c}}-\phi ^{\prime }(\eta _c)=0\,,\quad \text{ or\quad }%
\eta _c\left( \phi ^{\prime }(\eta _c)\right) ^2=-s\,.  \label{spe}
\end{equation}

If we define: 
\begin{equation}
\chi \left( s\right) =\phi \left( \eta _c\right) -2i\sqrt{s\eta _c}=\phi
\left( \eta _c\right) -2\eta _c\phi ^{\prime }(\eta _c)\,,  \label{spt}
\end{equation}
the following relations can be easily established: 
\begin{eqnarray*}
\chi ^{\prime }(s)=\frac 1{\phi ^{\prime }(\eta _c)}\,,\quad s\left( \chi
^{\prime }(s)\right) ^2= &-\eta _c\,,\quad \phi (\eta _c)=&\chi (s)-2s\chi
^{\prime }(s)\,,\quad \\
\chi ^{\prime }(s)+2s\chi ^{\prime \prime }(s) &=&\frac 1{\phi ^{\prime
}(\eta _c)+2\eta _c\phi ^{\prime \prime }(\eta _c)}\,.
\end{eqnarray*}
Obviously, the transformation $\phi (\eta )\leftrightarrow \chi (s)$ is
symmetric, i.e., its reverse has the same functional form. With $\chi(s)$ we
may rewrite Eq.(\ref{sp}) as: 
\begin{equation}
Q(s)=\left( 1-c\right) \left[ 1+2s\frac{\chi ^{\prime \prime }(s)}{\chi
^{\prime }(s)}\right] ^{1/2}\exp \left( -\chi (s)\right) .  \label{sp1}
\end{equation}

Proceeding along the same line, one also can derive the following equality: 
\begin{eqnarray}
&&\sqrt{\eta }\int_0^\infty \frac{ds}{\sqrt{s}}J_1\left( 2\sqrt{s\eta }
\right) Q(s)\ln Q(s)=\left( 1-c\right) \ln \left( 1-c\right) +  \nonumber \\
&&\left( 1-c\right) \left[ \chi \left( s_c\right) -\ln \left( 1-c\right) - 
\frac 12\ln \left( 1+2s_c\frac{\chi ^{\prime \prime }(s_c)}{\chi ^{\prime
}(s_c)}\right) \right] \exp \left( -\chi (s_c)+2s_c\chi ^{\prime
}(s_c)\right) =  \label{sp2} \\
&&\left( 1-c\right) \left[ \phi (\eta )-2\eta \phi ^{\prime }(\eta )+ \frac 1%
2\ln \left( 1+2\frac{\phi ^{\prime \prime }(\eta )}{\phi ^{\prime }(\eta )}%
\right) \right] \exp \left( -\phi (\eta )\right).  \nonumber
\end{eqnarray}
Replacing $\phi (\eta )=-\ln \left[P(\eta )-c\right]$ back in Eq. (\ref{sp2}%
), and substituting the result into Eq. (\ref{eveq}), we get the evolution
equation (\ref{eveq}) in the differential form: 
\begin{eqnarray}
\lambda \frac{\partial P}{\partial \lambda } &=&B_1\left( \left\{ P\right\}
,\eta \right) =-\left( 1+\epsilon \right) \eta P^{\prime }+\epsilon P\ln
P-\left( 1-c\right) \ln \left( 1-c\right) +\left( P-c\right) \ln \left(
P-c\right)  \nonumber \\
&-&\frac 12\left( P-c\right) \ln \left[ 1-2\eta \frac{P^{\prime \prime }}{
P^{\prime }}-2\eta \frac{P^{\prime }}{P-c}\right] \,.  \label{evsp}
\end{eqnarray}

The derivation of Eq. (\ref{evsp}) suggests the replacement of the evolution
operator $B$ with its approximate form $B_1$ at least for large enough $\eta$%
. But, if the function $P(\eta )$ is analytic, then, taking into account $%
P(0)=1$, it may be represented in the form (\ref{lin}) in some neighborhood
of the point $\eta =0$. If we plug Eq. (\ref{lin}) into Eq. (\ref{evsp}),
and linearize the resulting expression with respect to $\psi $, we arrive 
{\em exactly} at the same evolution operator as in Eq. (\ref{linev}), which
was obtained by the linearization of the exact evolution operator $B $. This
observation prompts us to enhance the region of validity of the approximate
evolution equation (\ref{evsp}) for everywhere in the complex plane $\eta $.

At the percolation threshold, $c=c_t$, the solution of the RG evolution
equation can be taken in the form: 
\begin{eqnarray}
P(\eta,\lambda )=\bar{P}(\eta\lambda ^{-a}).  \label{anz1}
\end{eqnarray}
Here the critical index $a$ is related to the critical exponents of the
conductivity $\mu$ and of the correlation length $\nu$ by the relation 
\begin{eqnarray}
a=\mu /\nu.  \label{def:a}
\end{eqnarray}
Indeed, according to Eqs. (\ref{def1},\ref{anz1}), the average conductivity
of the $\lambda$-sized cube in the critical regime, $c=c_t$, is 
\begin{eqnarray}
<\sigma(\lambda)> = {\frac{dP(\eta,\lambda)}{d\eta}}|_{\eta=0}
=\sigma_0\left(\frac{\lambda_0}{\lambda}\right)^a.  \label{17.28b}
\end{eqnarray}
The same conductivity (\ref{17.28b}) is realized in the infinite disordered
fractal with the correlation length $\xi$ equal to $\lambda$. Near the
percolation threshold the correlation length $\xi$ is given by Eq. (\ref
{defnu}) and, therefore, according to Eq. (\ref{17.28b}) the fractal
conductivity obeys the scaling law 
\begin{eqnarray}
<\sigma> \sim (c_t-c)^\mu,~~~~~~\mu = a\nu.  \nonumber
\end{eqnarray}

With the scaling anzats (\ref{anz1}), Eq. (\ref{evsp}) becomes an ordinary
differential one of the second order. It appears to be more convenient to
use the function $\phi (x)=-\ln \left[ \left( \bar{P }(x)-c_t\right)
/(1-c_t)\right] $ instead of $\bar{P}(x)$. Denoting $\phi _0=-\ln \left[
c_t/(1-c_t)\right] $, we have: 
\begin{equation}
\frac 12\ln \left[ 1+2x\frac{\phi ^{\prime \prime }}{\phi ^{\prime }}\right]
=\left( 1+\epsilon -a\right) x\phi ^{\prime }-\phi +\epsilon \left[ g\left(
\phi -\phi _0\right) -g\left( -\phi _0\right) \right] \,,  \label{steq}
\end{equation}
where we introduce: 
\begin{equation}
g(\phi )\equiv \left( e^\phi +1\right) \ln \left( 1+e^\phi \right) \,.
\label{defg}
\end{equation}
An equation for $\phi _0$ which follows from Eq. (\ref{thrp}) was used to
derive Eq. (\ref{steq}). The latter may be reduced to the first order linear
equation by introducing: 
\begin{equation}
z\left( \phi \right) \equiv \exp \left[ -2\left( 1+\epsilon -a\right) x\phi
^{\prime }\right] \,,  \label{defz}
\end{equation}
and treating $\phi $ as independent variable: 
\begin{equation}
\frac{dz}{d\phi }+\left( 1+\epsilon -a\right) z=-\left( 1+\epsilon -a\right)
\exp \left\{ -2\phi +2\epsilon \left[ g\left( \phi -\phi _0\right) -g\left(
-\phi _0\right) \right] \right\} \,,  \label{steq1}
\end{equation}
where $a$ is related to the critical index of conductivity by Eq. (\ref
{def:a}).

One should require the solution of Eq. (\ref{steq1}) $z(\phi )\rightarrow 0$
as $\phi \rightarrow +\infty $ faster than $\exp \left[ -\left( 1+\epsilon
-a\right) \phi \right] $ to ensure the applicability of the saddle-point
approximation. This selects a solution in the form 
\begin{equation}
z\left( \phi \right) =\left( 1+\epsilon -a\right) e^{-\left( 1+\epsilon
-a\right) \phi }\int_\phi ^\infty dy\exp \left\{ -\left( 1+\epsilon
-a\right) y+2\epsilon \left[ g\left( y-\phi _0\right) -g\left( -\phi
_0\right) \right] \right\}.  \label{soln1}
\end{equation}
The normalization condition $\phi (0)=0$ implies $z(0)=1$, from which the
equation for $a$ follows: 
\begin{equation}
\left( 1+\epsilon -a\right) \int_0^\infty dy\exp \left\{ -\left( 1+\epsilon
-a\right) y+2\epsilon \left[ g\left( y-\phi _0\right) -g\left( -\phi
_0\right) \right] \right\} =1\,.  \label{eqa}
\end{equation}

Comparing the values of $a$, obtained from Eq.(\ref{eqa}) and by the
numerical investigation of the original evolution equation (\ref{eveq}), one
can see that both methods give the same results at any dimensionality. This,
together with the considerations presented above, prompts us to consider the
saddle point solution as an exact one. Of course, the RGMK method itself is
an approximate one.

Comparison of the numeric results for the critical exponent of conductivity $%
\mu $ is presented in Table \ref{exponents}. In three dimensions we have
from Eq. (\ref{eqa}): $a\approx 1.891$. On the other hand, the best
numerical results \cite{nh97} give $a=2.25\pm 0.04$. So, the RGMK method
provides a reasonable solution even for 3d systems.

For the case $\epsilon \ll 1$, from Eq. (\ref{eqa}) it follows 
\begin{equation}
a=\frac{1+\epsilon }\epsilon \exp \left( -\frac{1+\epsilon }\epsilon \right)
\,,\;\;\;\mu =\frac a\epsilon.  \label{cexld}
\end{equation}
For $\epsilon \gg 1$ from Eq. (\ref{eqa}) one may obtain: 
\begin{equation}
a=\epsilon -\frac \epsilon 4e^{-\epsilon }\,.  \label{cexhd}
\end{equation}

\section{Distribution function at the threshold}

\label{DF}

The function $\phi (x)$ may be determined as a reverse of the equation: 
\begin{equation}
Cx=\phi \exp \left[ -\int_0^\phi d\zeta \left( \frac{1+\epsilon -a}{z(\zeta
) }-\frac 1\zeta \right) \right] \,.  \label{sol2}
\end{equation}
The integration constant $C$ in Eq. (\ref{sol2}) corresponds to arbitrary
choice of the unit of conductivity, or, alternatively, of the length scale
at the percolation threshold.

Thus the distribution function (DF) for conductivities which in the initial
representation is defined as, $\Pi (\sigma,\lambda )=\left\langle \delta
\left( \sigma_\lambda -\sigma \right) \right\rangle $ at the percolation
threshold takes the universal scaling form: 
\begin{eqnarray}
\Pi (\sigma ,\lambda )=c_t+(1-c_t)\bar{\Pi}(y),  \label{17.45}
\end{eqnarray}
where $y$ is the conductivity in units of the average conductivity (\ref
{17.28b}) 
\begin{eqnarray}
y={\frac{\sigma }{<\sigma(\lambda)>}}={\frac{\sigma}{\sigma_0}}\left({\frac{%
\lambda}{\lambda_0}}\right)^a,  \label{17.37a}
\end{eqnarray}
and the scaling function $\bar{\Pi}(y)$ may be expressed as the integral: 
\begin{equation}
\bar{\Pi}(y)=\int_{-i\infty }^{+i\infty }\frac{dx}{2\pi i}\exp \left[
xy-\phi (x)\right] =\frac 1y\int_{0-i\infty }^{0+i\infty }\frac{d\phi }{2\pi
i} \exp \left[ -\phi +yx(\phi )\right].  \label{scf}
\end{equation}
The last equality was obtained through integration by parts.

However, it should be noted that additional unphysical contribution arises
when evaluating the integral in Eq. (\ref{scf}). Namely, the function $%
x(\phi )$ is singular at $\phi =\tilde{\phi}_n=\phi _0+i\pi (2n+1)$, where $%
n $ is an integer number. These singularities result from the procedure of
analytic continuation within the RGMK approach. It can be illustrated as
follows: Let us assume the initial distribution of conductivities to be: $%
P_0(\eta )=c+(1-c)e^{-\eta }$. After putting $m$ identically distributed
conductors in parallel, the Laplace transform of DF for their sum, $P_1(\eta
)=\left[ c+(1-c)e^{-\eta }\right] ^m\,$, has $m$-th order zeroes at $\tilde{%
\eta}_n=\eta _0+i\pi (2n+1)$, where $\eta _0=\ln [(1-c)/c]$. These zeroes
transform into singularities after analytic continuation to non-integer $m$.
Thus the procedure of the transition from integer rescaling transformation
(which is exact for a hierarchical structure) to the infinitesimal one is
the source of the above singularities in Eq. (\ref{scf}). Since these
singularities are artificial ones they should be merely discarded in the
integral (\ref{scf}).

At large conductivities $y \gg 1$, shifting integration contour in Eq. (\ref
{scf}) to the region $\Re \phi >\phi _0$, one has the following asymptotic
expression for the DF: 
\begin{equation}
\bar \Pi (y\gg1)=\frac{D_2}{y_2}\left( \frac y{y_2}\right) ^{\frac{%
1+\epsilon }{ 2a}-1}\exp \left[ -\left( \frac y{y_2}\right) ^{\frac{%
1+\epsilon }a}\right] \,,  \label{leftscf}
\end{equation}
where: 
\begin{eqnarray}
D_2 &=&\frac{[2\pi (1+\epsilon )(1+\epsilon -a)]^{1/2}}{(1-c_t)a}\left( 
\frac{1+\epsilon +a}{1+\epsilon -a}\right) ^{\frac 1{2(1+\epsilon )}}\,,\ \
y_2=\frac{1+\epsilon }{1+\epsilon -a}\left( \frac{1+\epsilon -a}a\right) ^{%
\frac a{1+\epsilon }}e^{A_2}\,,  \nonumber \\
A_2 &=&2(1+\epsilon -a)\int_{-\infty }^0d\zeta \ln (-\zeta )\frac d{d\zeta } 
\frac \zeta {\ln z(\zeta )}.  \label{leftpar}
\end{eqnarray}

Shifting the integration contour in Eq.(\ref{scf}) to the region $\Re \phi
<\phi _0$, we arrive at the following expression for $\bar{\Pi}(y)$ in the
region of small $y$: 
\begin{equation}
\bar{\Pi}(y\ll1)=\frac{D_1}{y_1}\left( \frac y{y_1}\right) ^{\frac 1{%
2(\epsilon -a)}+1}\exp \left[ -\left( \frac{y_1}y\right) ^{\frac 1{\epsilon
-a}}\right] \,,  \label{rightscf}
\end{equation}
with: 
\begin{eqnarray}
D_1 &=&\frac{[2\pi (1+\epsilon -a)]^{1/2}}{\epsilon -a}e^{-\epsilon
}c_t^{-\epsilon /(1-c_t)}\,,\ \ y_1=\frac{e^{-A_1}}{1+\epsilon -a}\left( 
\frac{\epsilon -a}{1+\epsilon -a}\right) ^{\epsilon -a},  \nonumber \\
A_1 &=&2(1+\epsilon -a)\int_0^\infty d\zeta \ln \zeta \frac d{d\zeta }\frac %
\zeta {\ln z(\zeta )}\,.  \label{rightpar}
\end{eqnarray}

More detailed results are available in the limit $\epsilon \ll 1$. At $\Re
\phi <\phi _0\simeq 1/\epsilon $ to first order of $c_t$ and of $a$ we get
the following expression for $x(\phi )$: 
\begin{equation}
\ln x(\phi )=\ln \phi +c_t\frac{e^\phi -1-\phi }\phi +\frac a{(1+\epsilon
)^2 }\int_0^\phi \frac{d\zeta }{\zeta ^2}\left[ e^{(1+\epsilon )\zeta
}-1-(1+\epsilon )\zeta \right] \,.  \label{leftx1d}
\end{equation}
Evaluating Taylor's series of $\phi \left( x\right) $ at $x=0$, the central
momenta of conductivity are found to be of the order of $a$: 
\begin{equation}
\frac{\left\langle \sigma -\left\langle \sigma \right\rangle \right\rangle ^2%
}{\left\langle \sigma \right\rangle ^2}=a+c_t\,,\ \ \ \frac{\left\langle
\sigma -\left\langle \sigma \right\rangle \right\rangle ^3}{\left\langle
\sigma \right\rangle ^3}=-\frac 12(1+\epsilon )a+c_t\,,\ \ldots \ 
\label{mom1d}
\end{equation}

\begin{figure}[tbp]
\epsfbox{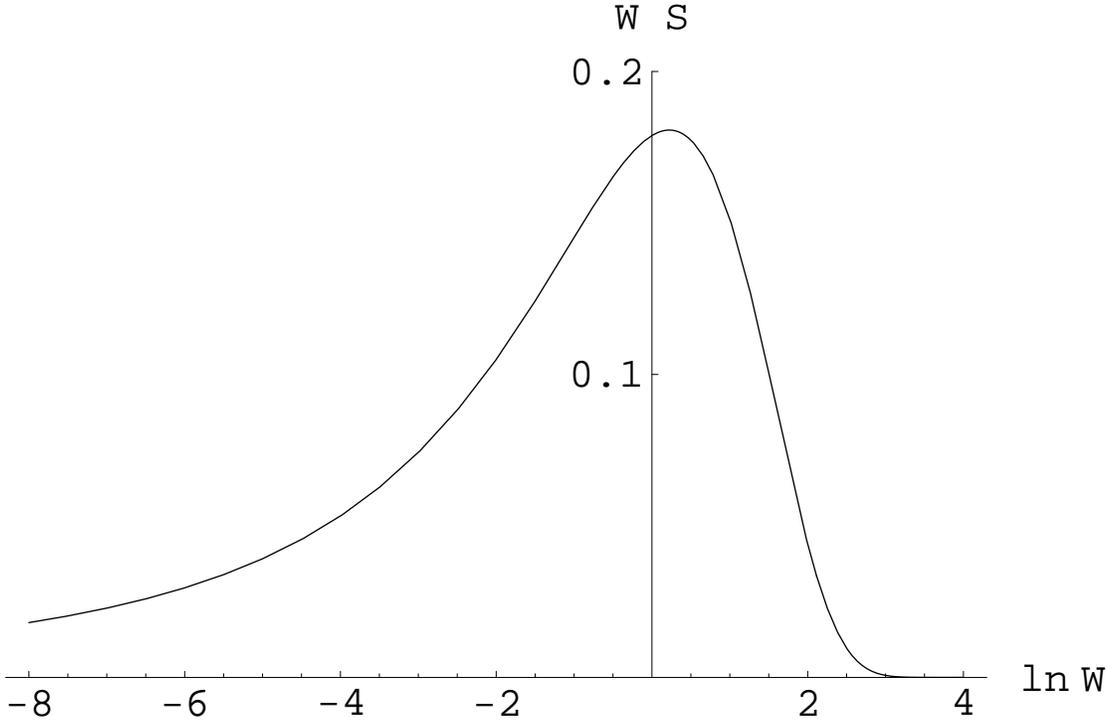}
\vspace{0.5cm}
\caption{Universal distribution of conductivity for the quasi-$1D$ fractal
at the percolation threshold (Eq. (\ref{a1daslarge})). There is a sharp
decay in the region of large conductivity and a long tail for large
resistances (Eq. (\ref{sfuncas})).}
\label{univ}
\end{figure}

On the other hand, using in Eq.(\ref{scf}) the asymptotics of $x(\phi )$ at $%
\Re \phi <\phi _0$ and $\left| \phi \right| \gg 1$ we have for large enough $%
y$ (see also Appendix \ref{apc}): 
\begin{equation}
y\bar{\Pi}(y)=e^{\phi _2}\frac{1+\epsilon }aWS(W)\,,  \label{a1daslarge}
\end{equation}
where $\phi _2={\frac{a}{(1+\epsilon )^2}}-c_t$, and the new fluctuating
variable was introduced: 
\begin{equation}
W=e^{G_1}\frac a{1+\epsilon }y^{\frac{1+\epsilon }a}\,,\ \ \ G_1=1-\gamma
-\ln (1+\epsilon )-\frac{(1+\epsilon )c_t}a\simeq 1-\gamma \,.
\label{Omegay}
\end{equation}
$\gamma $ is Euler's constant, and $S(W)$ is given by: 
\begin{equation}
S(W)=\int_{-i\infty +\Delta }^{i\infty +\Delta }\frac{du}{2\pi i}u^{Wu}\,.
\label{sfunc}
\end{equation}
An asymptotic expression for $S(W)$ may be easily obtained by the
saddle-point method (Appendix \ref{apc}): 
\begin{equation}
S(W)\approx \left\{ 
\begin{array}{l}
\frac{\exp (-e^{-1}W)}{\sqrt{2\pi eW}},\text{~as }W\gg 1\,; \\ 
\frac{\sqrt{2\pi }}{eW}\left[ \frac{\ln \ln (e/W)}{\ln (e/W)}\right] ^2, 
\text{~as }W\ll 1.
\end{array}
\right.  \label{sfuncas}
\end{equation}
Fig. \ref{univ} shows $WS\left( W\right) $ as a function of $\ln W$.

The asymptote of $\bar{\Pi}(y)$ at small $y$ is given by Eqs. (\ref{rightscf}%
, \ref{rightpar}). At $\epsilon \ll 1$ it may be reduced to: 
\begin{equation}
y\bar{\Pi}(y)=\frac{y^{-\frac 1{2\epsilon }}}{\sqrt{2\pi }\epsilon }\exp
\left( \frac 12-e^{-1}y^{-1/\epsilon }\right) \,.  \label{left1das}
\end{equation}
The two expressions (\ref{a1daslarge},\ref{left1das}) should be supplemented
by one for the intermediate region, where the function $x(\phi)$ in the
integral (\ref{scf}) can be expanded in $a$ and $c_t$. Here we have in the
first order of $a,\,c_t$: 
\begin{eqnarray*}
x\left( \phi \right) &=&\phi +\eta \left( \phi \right) ,\;\eta \left( \phi
\right) =c_t\left( e^\phi -1-\phi \right) +\frac{a\phi }{\left( 1+\epsilon
\right) ^2}\int_0^\phi \frac{d\zeta }{\zeta ^2}\left[ e^{\left( 1+\epsilon
\right) \zeta }-1-\left( 1+\epsilon \right) \zeta \right] \,, \\
y\bar{\Pi}\left( y\right) &=&\delta \left( y-1\right) +y\int_{-i\infty
}^{+i\infty }\frac{d\phi }{2\pi i}e^{\left( y-1\right) \phi }\eta \left(
\phi \right) \,.
\end{eqnarray*}
Apart from the $\delta $-function term, this yields in the region $0<y<1$ : 
\begin{equation}
\bar{\Pi}(y)=\frac a{1+\epsilon }\frac 1{\Delta ^2}\,,  \label{a1dinm}
\end{equation}
where $\Delta =1-y$.

To establish regions of validity for different expressions of DF, let us
consider the region $\Delta \ll 1$. Here Eq.(\ref{left1das}) turns into: 
\begin{equation}
\bar{\Pi}\left( y\right) =\frac 1{\sqrt{2\pi }\epsilon }\exp \left( \frac 12+%
\frac \Delta {2\epsilon }-e^{\Delta /\varepsilon -1}\right) \,,
\label{left1das1}
\end{equation}
and Eq.(\ref{a1daslarge}) may be written, taking into account Eq.(\ref
{sfuncas}) at $W\ll 1$, as: 
\begin{equation}
\bar{\Pi}\left( y\right) =\frac{\sqrt{2\pi }}e\frac a{1+\epsilon }\left[ 
\frac{\ln \left( \ln \frac{1+\epsilon }a+\frac{1+\epsilon }a\Delta \right) }{%
\Delta +\frac a{1+\epsilon }\ln \frac{1+\epsilon }a}\right] ^2.
\label{a1daslarge1}
\end{equation}
Comparing Eq.(\ref{a1dinm}) with Eqs.(\ref{a1daslarge1},\ref{left1das}) one
can conclude that Eq.(\ref{a1daslarge}) is valid if $\Delta <\Delta _1\sim
a/\epsilon \ln \left( 1/\epsilon \right) $, and Eq.(\ref{left1das}) holds
for $\Delta >\Delta _2\sim \epsilon \ln \left( e/\epsilon \right) $.
Fluctuations of conductivity appears to be distributed within narrow region
of relative width $\Delta _1$, which ensures that not very high order
central momenta of the conductivity to be small (see Eq.(\ref{mom1d})).
However, if expressed in terms of the universally fluctuating variable $W $,
the distribution becomes smeared over a wide region with the lower cut-off $%
W_1\sim a^p$, $p\approx 1+1/\ln \left( 1/\epsilon \right) $.

The distribution function $S\left( W\right)$ arises naturally in a 1d chain
of random resistors, if, to require a scaling form for the distribution
function of $\lambda $-length chain: $\Upsilon \left( \rho,\lambda \right) =%
\bar{\Upsilon}(\rho\lambda ^{-a})$, or $Q\left( s,\lambda \right) = \bar{Q}%
\left( s\lambda ^a\right) $ in the Laplace representation. Then from $%
Q\left( s,n\lambda \right) =Q^n\left( s/n,\lambda \right) $ it immediately
follows: $\bar{Q} \left( x\right) =\exp \left[ -Cx^{1/\left( 1+a\right)
}\right] $. Evaluating its inverse Laplace's transform $\Upsilon \left(
r\right) $, and assuming $a\ll 1$, which is true in the 1d case, we have
after the proper rescaling of the integration variable: 
\begin{equation}
r\Upsilon \left( r\right) =\frac 1aWS\left( W\right) ,\;\;W=ar^{-1/a},
\label{scf1d}
\end{equation}
which is essentially the same formula as Eqs. (\ref{a1daslarge},\ref{Omegay}%
).

\section{Scaling and AC conductivity}

\label{scale}

Exact results for AC-conductivity in disordered systems are available for a
very limited class of models, mostly for 1d ones. The common method to study
disordered hopping systems is the effective medium approximation (EMA),
which gives qualitatively correct results for three-, two-, and even for
one-dimensional systems. However, it fails for a nearly 1d system. In a
percolation model, for an example, EMA gives threshold concentration value $%
c_t\propto \epsilon $ and completely wrong values of critical exponents.
However, knowing the results for DC conductivity and topological properties
of the percolation network, the qualitative behavior of low-frequency
conductivity may be restored within the scaling hypothesis \cite{nyo94,sw94}.

Namely, it should be assumed that the only length scale near the threshold
is the correlation length $\xi \propto |\tau |^{-\nu }$, $\tau =(c_t-c)/c_t$%
. The second assumption is about the anomalous diffusion of a tracer placed
onto the infinite cluster at the percolation threshold: 
\begin{equation}
\left\langle r^2(t)\right\rangle _\infty \propto t^\zeta \,,\quad \zeta <1\,.
\label{adif}
\end{equation}
$r(t)$ is the distance from tracer's position at $t=0$, $\left\langle \dots
\right\rangle _\infty $ means the average over the initial positions within
the infinite cluster only. Above the threshold, when $c<c_t$, we have normal
diffusion at sufficiently large times, when $\left\langle
r^2(t)\right\rangle >\xi ^2$: 
\begin{equation}
\left\langle r^2(t)\right\rangle =D_\infty t\,,  \label{ndif}
\end{equation}
with the diffusion constant $D_\infty $ connected with DC conductivity as: 
\begin{equation}
\sigma _{DC}=\frac{e^2n_e}{kT}D_\infty \,.  \label{dc}
\end{equation}
$e$ and $n_e$ are charge and concentration of electrons, respectively. Near
threshold we have $D_\infty \propto \tau ^\mu $, $\mu $ is the critical
exponent of the DC conductivity. Note, however, that in Eq. (\ref{ndif}) the
average is over the whole network, including finite clusters.

The relationship between exponents $\mu $ and $\zeta $ may be established in
the following way: At some large enough time $t$ let us consider the $%
\lambda $-sized box, $\lambda =\left\langle r^2(t)\right\rangle
^{1/2}\propto t^{\zeta /2}$ inside the system at the percolation threshold.
Its conductivity is $\sigma (\lambda )\propto \lambda ^{-\mu /\nu }$. At the
same time, it may be expressed as $\sigma (\lambda )\propto P_\infty
(\lambda )\lambda ^2/t$, where $P_\infty (\lambda )\equiv P_1$ is the
infinite cluster capacity for the system in which the correlation length
equals $\lambda $. Since $P_\infty (\lambda )\propto \lambda ^{-\beta /\nu }$%
, it follows $\sigma (\lambda )\propto \lambda ^{\beta +2-2/\zeta }$. Thus
we can conclude: 
\begin{equation}
\zeta =\frac{2\nu }{2\nu +\mu -\beta }\,\,.  \label{dzmu}
\end{equation}
If we take into account the average over finite clusters, we obtain at the
percolation threshold: $\left\langle r^2(t)\right\rangle \propto P_\infty
(\lambda )\lambda ^2\propto \lambda ^{2-\beta /\nu }\propto t^{\bar{\zeta}}$
, $\bar{\zeta}$ is the anomalous diffusion exponent including the
contribution of finite clusters: 
\begin{equation}
\bar{\zeta}=\left( 1-\frac \beta {2\nu }\right) \zeta =\frac{2\nu -\beta }{
2\nu +\mu -\beta }\,.  \label{dzbar}
\end{equation}
Below the percolation threshold, at $c>c_t$, we have $\left\langle
r^2(t)\right\rangle \sim \xi ^2\propto \tau ^{-2\nu }$ at $t\rightarrow
\infty $.

All the above may be summarized as: 
\begin{equation}
\left\langle r^2(t)\right\rangle =t^{\bar \zeta }G(\tau t^{u/\mu })\,,
\label{scal}
\end{equation}
where: 
\begin{equation}
u=1-\bar \zeta =\frac \mu {2\nu +\mu -\beta }\,,  \label{frexp}
\end{equation}
with the scaling function $G(0)$ being some constant, $G(x)\propto x^\mu $
as $x\rightarrow +\infty $, and $G(x)\propto |x|^{-s}$, 
\begin{equation}
s=2\nu -\beta \,,  \label{diexp}
\end{equation}
as $x\rightarrow -\infty $.

The conductivity may be expressed through $\left\langle r^2(S)\right\rangle $%
, the Laplace transform of $\left\langle r^2(t)\right\rangle $, as: 
\begin{equation}
\sigma \left( \omega ,\tau \right) =\frac{e^2n_e}{kT}S^2\left\langle
r^2(S)\right\rangle \,,  \label{acc}
\end{equation}
where $S=-i\omega $. Using the Tauberian theorem for the Laplace
transformation of power laws, from Eqs. (\ref{scal},\ref{acc}) one can
obtain: 
\begin{equation}
\sigma (\omega ,\tau )=\frac{e^2n_e}{kT}S^u\bar G\left( \tau S^{-u/\mu
}\right) \,,  \label{scal1}
\end{equation}
with scaling function $\bar G(x)$ having the same asymptotic properties as $%
G(x)$.

\begin{figure}[tbp]
\epsfbox{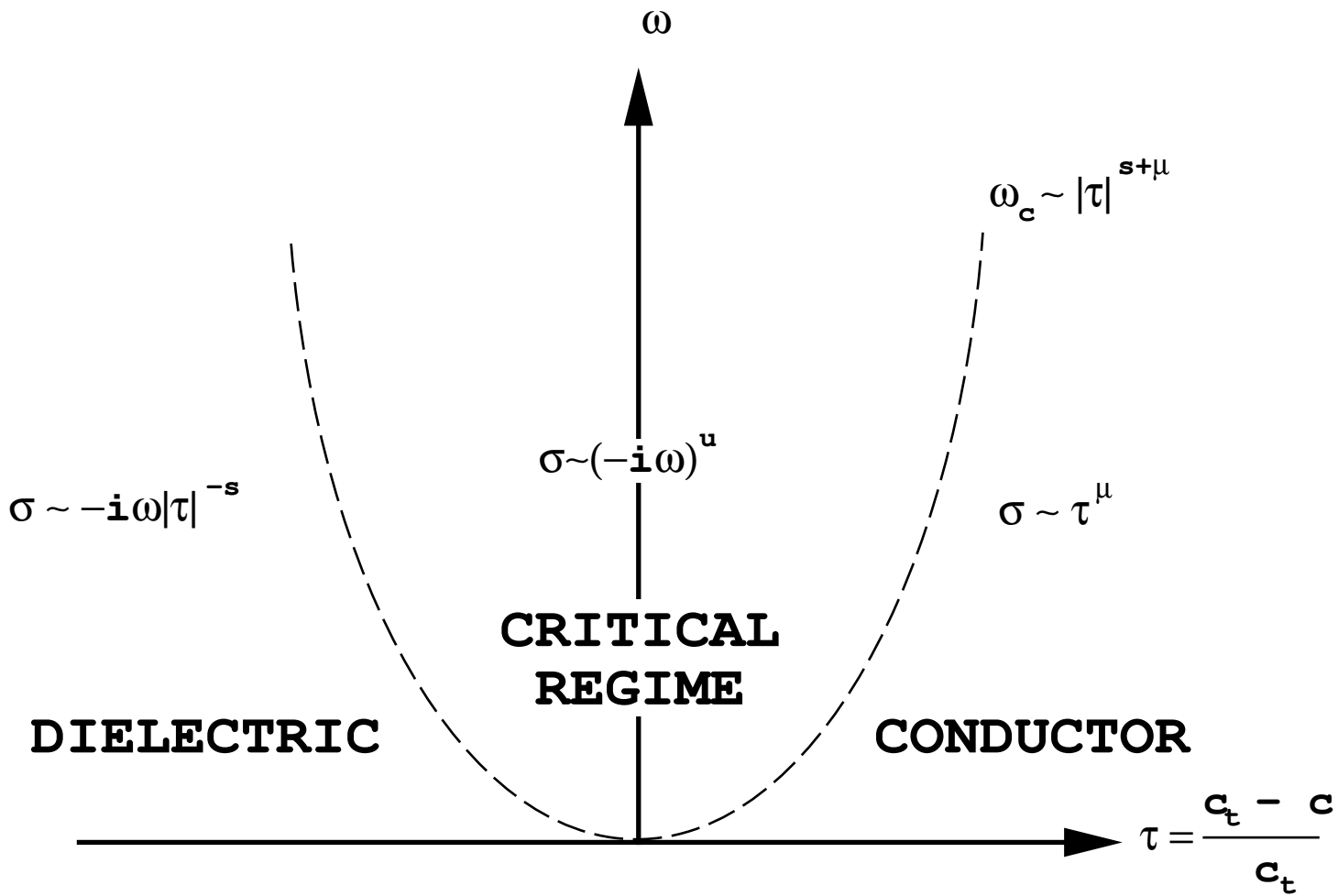}
\vspace{6.5cm}
\caption{ Diagram of frequency-dependent conductivity in the quasi-$1D$
fractal near the percolation threshold, $|\tau|\ll 1$. Here $s=2\nu-\beta$
and $u = \mu/(2\nu+\mu-\beta)$ and $\nu$, $\beta$ and $\mu$ are the critical
indexes of the correlation length, capacity of infinite cluster and
conductivity; $\omega_c \sim |\tau|^{s+\mu}$ is the boundary frequency
separated the critical region from the conducting and dielectric phases. For
the quasi-$1D$ fractal with the transverse dimensionality $\epsilon \ll 1$, $%
u \ll 1$, $\mu \ll 1$ and $s \gg 1$.}
\label{accond}
\end{figure}

Thus we have at $c<c_t$ and $\omega \ll \tau ^{s+\mu }$: 
\[
\sigma \propto \tau ^\mu, 
\]
at $c>c_t$ and $\omega \ll |\tau |^{s+\mu }$: 
\[
\sigma \propto -i\omega \tau ^{-s}, 
\]
and at $\omega \gg |\tau |^{s+\mu }$ (in particular, at any $\omega $ if $%
c=c_t$): 
\[
\sigma \propto (-i\omega )^u. 
\]
The summary of the frequency dependence of AC conductivity is given by Fig. 
\ref{accond}.

In a nearly 1d case the static conductivity exponent $\mu $ is given by Eq. (%
\ref{cexld}), the frequency dependence exponent $u$ and the exponent $s$ of
dielectric constant divergence may be written with Eqs. (\ref{clen1d},\ref
{beta}) as: 
\begin{equation}
s\approx \frac 2\epsilon \,,  \label{sap}
\end{equation}
\begin{equation}
u\approx \frac 1{2\epsilon }\exp \left( -1-\frac 1\epsilon \right) \,.
\label{uexp}
\end{equation}
Thus the exponent $s$ is very large but the exponent $u$ is very small.
Therefore, in a dielectric phase the AC conductivity as a function of
frequency demonstrates a step-type behavior. In the conducting state the
frequency dependence of conductivity remains very weak.

\section{Variable range hopping}

\label{vrh}

In a conductor with localized carriers the charge transport is provided by
the variable range hopping (VRH). The model may be formulated as follows.
The phonon assisted hopping rate $w_{ij}$ from one localized state $j$ to
the other $i$ per unit time, including Fermi occupation probabilities $p_i$,
is approximated by the formula \cite{kk73}: 
\begin{eqnarray}
w_{ij}p_j\left( 1-p_i\right) &=&\omega _0\exp \left( -2f_{ij}\right) \,, 
\nonumber \\
f_{ij} &=&\frac{\left| \varepsilon _j-\varepsilon _i\right| +\left|
\varepsilon _j - \varepsilon_F\right| +\left| \varepsilon _i
-\varepsilon_F\right| }{4kT}+\frac{\left| {\bf r}_j-{\bf r}_i\right| }a\,,
\label{hr}
\end{eqnarray}
where $\varepsilon _i$ and ${\bf r}_i$ are energies and position vectors of
localized states respectively, and $a$ is their radius. We assume here the
hopping motion to be along the chains. Assuming that the localized states
near the Fermi level are distributed uniformly in space and energy, the
distribution of random variables $f_{ij}$ is: 
\begin{equation}
F(f)\equiv \text{Probability}(f_{ij}>f)=\exp \left[ -\left( \frac f{f_0}
\right) ^2\right] \,,  \label{df}
\end{equation}
where: 
\begin{equation}
f_0=\left( \frac{T_0}T\right) ^{1/2}\,,\quad kT_0={\frac{1 }{4N_Fa}}\,.
\label{f0}
\end{equation}
$N_F$ is the density of states at the Fermi level.

To study charge motion in a system with continuous distribution of hopping
rates is a much more complicated problem than the one for a percolating
system, where hopping rates are either $0$, or some given finite value $w_0$%
. However, knowing the results for the conductivity of percolating system,
qualitative conclusions can be obtained for the system with continuously
distributed hopping rates. Namely, let us introduce some probe hopping rate $%
w_c$, replacing all hopping rates $w_{ij}<w_c$ with $0$, and all $w_{ij}>w_c$
with $w_c$. Obviously the conductivity becomes lower than the initial one,
but if to choose $w_c$ from the requirement to get a maximal conductivity,
one can hope to obtain a good estimate for the original conductivity.

It is convenient to represent the probe value as $w_c=\exp (-2f_c)$. The
corresponding broken bonds concentration $c$ is given by formula (\ref{df}),
i.e., $c=\exp \left[ -\left( f_c/f_0\right) ^2\right] $. The value of $f_c$
for the threshold concentration is $f_t=f_0/\sqrt{\epsilon }$ or 
\begin{eqnarray}
f_t={\frac{1}{2}}\left({\frac{T_1 }{T}}\right)^{1/2},~~~~~kT_1=\frac{ 4 kT_0}%
\epsilon =\frac 1{\epsilon N_Fa}.  \label{ft}
\end{eqnarray}
Assuming the probe value $f_c=f_t+\delta $ to be close to the threshold one, 
$\tau =\left( c_t-c\right) /c_t\ll 1$, we have the relation 
\begin{equation}
\delta =\frac 12\epsilon f_t\tau \,.  \label{dltau}
\end{equation}

The scaling formula (\ref{scal1}) for the conductivity of the percolation
system now reads 
\begin{equation}
\sigma (\omega ,\tau )=\frac{e^2n_e}{kT}a_{\parallel }^2|\tau |^\mu
w_cg\left[ \tau \left( \frac S{w_c}\right) ^{-u/\mu }\right] \,,
\label{scaling}
\end{equation}
where $w_c=w_t\exp \left( -\epsilon f_t\tau \right) $, $w_t=\omega _0\exp
\left( -2f_t\right) $, $a_{\parallel }=1/(N_FkT)$ is the hopping length, and
the electron density $n_e=N_FkT.$ Scaling function $g(x)$ has the following
properties: 
\[
g(x)\approx \left\{ 
\begin{array}{l}
A\left| x\right| ^{-\mu }\;\text{as }\left| x\right| \ll 1\,, \\ 
D+B_{+}x^{-s-\mu }+\ldots \text{ as }x\gg 1\,, \\ 
B_{-}\left| x\right| ^{-s-\mu }+\ldots \text{ as }x<0,\;|x|\gg 1,
\end{array}
\right. 
\]
with coefficients $A$, $B_{\pm}$ and $D$ of the order of unity.

First let us consider DC conductivity. Obviously one should choose $\tau >0$%
, and the expression to maximize the conductivity as a function of $\tau $
is $\tau ^\mu \exp \left( -\epsilon f_t\tau \right) $. The optimal value of
probe parameter $\tau $ is very small: $\tau _{DC}=\mu /\epsilon f_t$.
Taking into account Eq. (\ref{f0}), we have: 
\begin{eqnarray}
\sigma_{DC}\sim \frac{e^2}{N_F\left( kT\right) ^2}w_t,~~~~w_t=\omega _0\exp
\left[ -\left( \frac{T_1}T\right) ^{1/2}\right].  \label{dcc}
\end{eqnarray}
Thus the DC conductivity obeys a quasi-1d Mott's law, but the characteristic
temperature $T_1$ given by Eq. (\ref{ft}) is much greater than $T_0$ for VRH
in a strictly 1d chain from Eq. (\ref{f0}).

The so-called ``hydrodynamic region'' of very low frequencies, where the
conductivity's frequency dependence is determined by expansion: 
\begin{equation}
\sigma =\sigma _{DC}\left( 1-\frac{i\omega }{\omega _h}+\dots \right)
\,,\quad \omega <\omega _h\,,  \label{hydro}
\end{equation}
appears to be very narrow for the nearly 1d system. Its width $\omega _h$
may be estimated from the condition on the argument of the scaling function $%
g$ in Eq. (\ref{scaling}) to be of the order of unity at $\tau =\tau _{DC}$
and $\omega =\omega _h$: 
\begin{equation}
\omega _h\sim \tau _{DC}^{2/\epsilon }w_t\sim \exp \left( -\frac 2{\epsilon
^2}\right) f_t^{-2/\epsilon }w_t\,.  \label{oh}
\end{equation}
Within this region $|\omega|< \omega_h$ we get the effective value of $\tau
_c$ to be dependent on frequency as $\tau _c=\tau _{DC}\left( 1-S/\omega
_h+\dots \right) $. We suppose at further derivations $S=-i\omega $ to be
real and positive, having in mind analytic continuation afterwards.

At $S\sim \omega _h$, $\tau _c$ changes its sign, and now the conductivity
is determined by the charge motion inside finite-size clusters. At $|\tau
_c|\ll 1$, the size of effective clusters is large, i.e., the clusters
contain many 1d chains. This frequency region is called {\em multiple hopping%
} one. Note, that in contrast to two- and three-dimensional systems for
which the multiple hopping regime transforms at higher frequencies into the
regime of pair hops, here the multiple hopping frequency region borders that
of the one-dimensional hopping.

From properties of the scaling function $g$ in Eq. (\ref{scaling}) one can
conclude, that the conductivity is maximal if one chooses the probe value $%
\tau _c$ such that the argument of $g$ is of the order of unity. Taking into
account explicit expressions for critical exponents in a nearly 1d case we
have: 
\begin{equation}
\tau ^{\prime }\exp \left( \frac 12\epsilon ^2f_t\tau ^{\prime }\right) \sim
\left( \frac S{w_t}\right) ^{\epsilon /2}\,,  \label{mhe}
\end{equation}
where $\tau ^{\prime }\equiv -\tau _c$ and the conductivity may be estimated
to be: 
\begin{equation}
\sigma \sim \frac{e^2}{N_F(kT)^2}S\tau ^{\prime -2/\epsilon }\sim \frac{e^2}{
N_F(kT)^2}w_t\exp \left( \epsilon f_t\tau ^{\prime }\right) \,.  \label{mhc}
\end{equation}

One can see from Eq. (\ref{mhe}), that the character of frequency dependence
is determined by the parameter $\epsilon ^2f_t\equiv 2(T_2/T)^{1/2}$, where: 
\begin{equation}
T_2=\frac 14\epsilon ^4T_1\,.  \label{t2}
\end{equation}
If the temperature is relatively high, $T\gg T_2$, from Eq. (\ref{mhe}) it
follows $\tau ^{\prime }\sim \left( S/w_t\right) ^{\epsilon /2}$, and the
conductivity reads: 
\begin{equation}
\sigma \sim \frac{e^2}{N_F(kT)^2}w_t\exp \left[ \epsilon f_t\left( \frac S{
w_t}\right) ^{\epsilon /2}\right] \,.  \label{imf}
\end{equation}
Thus the temperature dependence of conductivity at a given frequency is
described by the quasi-1d Mott's law $\exp -\left( T_1/T\right) ^{1/2}$. The
frequency region for application of Eq. (\ref{imf}) is determined by the
requirement $\tau ^{\prime }\ll 1$, which may be written as: 
\begin{equation}
\frac \epsilon 2\ln \frac{w_t}S\gg 1\,.  \label{ub1}
\end{equation}

At lower temperatures, $T\ll T_2$, Eq. (\ref{imf}) is valid as long as $%
(1/2)\epsilon ^2f_t\tau ^{\prime }\ll 1$, or: 
\begin{equation}
\frac \epsilon 2\ln \frac{w_t}S\gg \ln \frac{\epsilon ^2f_t}2=\frac 12\ln 
\frac{T_2}T.  \label{ub2}
\end{equation}
At higher frequencies the solution of Eq. (\ref{mhe}) is given by the
equation 
\[
\tau ^{\prime }\sim \frac 1{\epsilon f_t}\ln \frac S{\omega _1}\,, 
\]
where $\omega _1=\left( 2/\epsilon ^2f_t\right) ^{2/\epsilon }w_t=\left(
T/T_2\right) ^{2/\epsilon }w_t$, and the conductivity now is: 
\begin{equation}
\sigma \sim \frac{e^2}{N_F(kT)^2}S\left( \frac{\epsilon f_t}{\ln \frac S{
\omega _1}}\right) ^{2/\epsilon }\,.  \label{hf}
\end{equation}
This formula remains to be valid until $\tau ^{\prime }\ll 1$, i.e. if: 
\begin{equation}
\frac \epsilon 2\ln \frac S{w_t}\ll \epsilon f_t\text{\thinspace .}
\label{upb}
\end{equation}

If the frequency is higher than ones determined by Eqs. (\ref{ub1}) or (\ref
{upb}), the conductivity behavior becomes a 1d one (see, e.g., Ref. \cite
{nps89}).

\section{Conclusions.}

\label{concl}

As one naturally expects, the results for the percolation problem in nearly
one dimension approach 1d ones as $\epsilon \rightarrow 0$. In particular,
threshold concentration $c_t$ (Eq. (\ref{thrn1d})) and critical exponents $%
\beta $ and $\mu$ (Eqs. (\ref{beta},\ref{cexld})) tend to zero. As a result
the capacity and conductivity have a jump-like behavior as a function of
concentration of broken bonds near a critical value. The reason is that in
the limit of one dimension the infinite cluster arises at $c=0$ and occupies
immediately the whole system. The critical length exponent $\nu \approx
\epsilon ^{-1}$ is, however, large, contrary to the 1d case, when $\nu =1$,
but this nearly 1d behavior of the correlation length can be observed in a
very narrow range of concentrations, $|c-c_t|\ll 1$. Outside this region,
when $c_t\ll c\ll 1$, critical length scales in a 1d manner, $\xi =c^{-1}$.

The other surprising feature is the strongly nonanalytic behavior of both
threshold concentration and of critical exponents, which points to a regular 
$\epsilon $-expansion near lower critical dimensionality $d=1$ being rather
impossible. Although the RGMK method becomes exact only in the limit $%
\epsilon \ll 1$ the comparison with numerical results (see Table 1) points
out that the critical indexes obtained by this method proves realistic even
for $\epsilon = 1,2$.

All these features together teach us that the infinite cluster arises almost
like a jump. The infinite cluster at the percolation threshold itself is a
fractal with a number of dimensionalities \cite{nyo94} all of them less than
the dimensionality of the original lattice itself. For example, fractal
dimensionality: 
\begin{equation}
D_f=D-\frac \beta \nu \approx 1+\epsilon -\frac 13\exp \left( -\frac 2%
\epsilon \right) \,,  \label{fd}
\end{equation}
characterizing the mass distribution within infinite cluster, is very close
to the fractal dimension of the system itself $D$, which means that infinite
cluster at the threshold is ``almost dense''.

Fracton, or spectral dimension $\tilde d$: 
\begin{equation}
\tilde d=D_f\frac{2\nu }{2\nu +\mu -\beta }\approx 1+\epsilon -\frac 1{%
2\epsilon }\exp \left( -1-\frac 1\epsilon \right) \,,  \label{fnd}
\end{equation}
was introduced to describe the behavior of random walk on the infinite
cluster (it may also be used to describe, e.g. density of localized
vibrational states, or fractons, etc.). Its closeness to $D=1+\epsilon $
means that the diffusion on the infinite cluster at the threshold is almost
normal. Respectively, the conductivity frequency dependence exponent $u$,
Eq. (\ref{uexp}), is small. On the other hand, dielectric constant in the
insulating phase, $\varepsilon ^{\prime }\propto |\tau |^{-s}$ diverges very
strongly, nearly as $\xi ^2\propto |\tau |^{-2\nu }$, but, similar to the
correlation length $\xi $, this divergency takes place in a narrow interval
of concentrations of the order of $c_t$.

The RGMK enables us to study not only the average characteristics of the
system but also their fluctuations. These fluctuations become essential near
the critical point when the correlation length $\xi$ becomes larger or
comparable with the system size $\lambda$. In this case because of
non-self-averaging a sample demonstrates individual characteristics
corresponding to its specific disorder.

In the present work we have found the distribution of possible
conductivities of samples in the critical regime. The average conductivity $%
<\sigma(\lambda)>$ decays with the sample size $\lambda$ according to
scaling law (\ref{17.28b}). All the fluctuations are found to obey the same
scaling law (\ref{mom1d}). Thus the distribution of conductivity is the
universal function, $\bar{\Pi}(y)$, in units of the average conductivity, $%
y=\sigma/<\sigma(\lambda)>$. In other words the above fractal
dimensionalities of the percolating cluster do not vary with the fractal
size, $\lambda$, as happens in multi-fractal systems \cite{Castellani86}.
This robustness comes from the additive laws (\ref{adlaw}) for classical
charge transport.

The function $\bar{\Pi}(y)$ represents the distribution of possible
experimental deviations from the scaling law (\ref{17.28b}). It is shown,
Eq. (\ref{mom1d}), that the central body of the distribution $\bar{\Pi}(y)$
is concentrated in the narrow interval around the average value, i.e., $y=1$%
, but it does not take the gaussian form. The distant tails of the
distribution in the region of large conductivity $y\gg 1$ and for large
resistivities $1/y \gg 1$ decay like a stretched exponent, Eqs. (\ref
{leftscf},\ref{rightscf}). In the limit $\epsilon \ll 1$ the shape of the
distribution function (\ref{a1daslarge}) is consistent with the 1d scaling
of the percolating cluster.

Returning to the variable range hopping in the chain fractal as a
consequence of (i) 1d character of variable range hopping along the chains,
and (ii) finite value of broken bonds threshold concentration $c_t$, the DC
conductivity obeys a quasi-1d Mott's law (\ref{dcc}). But the characteristic
temperature $T_1$ of this dependence is higher than the formal value of the
characteristic temperature for 1d chain (remembering that Mott's law is not
valid for 1d systems), $T_0$, by a factor $1/\epsilon$. This increase can be
understood through comparison with the quasi-1d model of weakly coupled
metallic chains \cite{nps89}. The variable range hopping conductivity of
this model obeys the same law with the characteristic temperature $T^*= {%
\frac{T_0}{2(d-1)}}$, where $2(d-1)$ is the number of neighboring chains and 
$d-1$ is the transverse dimensionality of the quasi-1d system ($d=3$).
Taking $d-1=\epsilon$ we formally reproduce $T_1$ for the nearly-1d fractal.
Experimental temperature dependence of conductivity in the poorly conducting
polymers very often follows a quasi-1d Mott's law with substantially
increased characteristic temperature \cite{wjrme90,joea94,joo98}, as
expected for $\epsilon \ll 1$.

In 2d and 3d isotropic systems with VRH mechanism of charge transport the
temperature dependencies of the AC conductivity and of the dielectric
constant are rather weak, but in nearly 1d systems there exists the region
of frequencies and temperatures, where these dependencies are nearly the
same as the quasi-1d Mott's type (see Eq. (\ref{imf})), which continuously
transforms into 1d dependence within a wide enough transient region, Eq. (%
\ref{hf}). Such a type of strong temperature dependence for both DC and AC
conductivity and also dielectric constant is experimentally observed in
conducting polymers with localized carriers \cite{wjrme90,joea94}.

The physical picture behind this dependence is the following. The
low-dimensional random system can be separated in weakly coupled clusters
within which carriers are confined. With increasing temperature, the size of
clusters exponentially increases, as more space accessible for carriers due
to thermal activation. As a result, the dielectric constant and the
conductivity exhibit strong temperature dependencies. In contrast to the
low-dimensional case, the clusters in two- and three-dimensional systems
prove to be more effectively coupled. Therefore the large polarization of
clusters does not happen because of transition of carriers between clusters.
Thus our results support strongly the idea that even poorly conducting
polymers represent low dimensional systems.

\acknowledgments

The authors are grateful to S.N. Dorogovtsev, V.V. Bryksin, Yu.A. Firsov, W.
Wonneberger and W. Schirmacher for useful discussions. This work was
partially supported by a Russian National Grants RFFI No. 96-02-16848 and
No. 97-02-18283, and the U.S. NSF DMR-9508723.

\appendix

\section{Evaluation of coefficients in the RG equation for ``free energy''.}

\label{apa}

Comparing Eqs. (\ref{trm}) and (\ref{ntrm}), we have: 
\begin{eqnarray}
f_q^{(0)}(c,h) &=&\ln T_{00}^{\prime }\,,\quad u(c,h)=\left. \frac \partial {%
\partial q}T_{00}^{\prime }\right| _{q=0}\,,\quad c^{\prime }=e^{-K^{\prime
}}=\frac{T_{l0}^{\prime 2}}{T_{00}^{\prime }T_{ll^{\prime }}^{\prime }}\,, 
\nonumber \\
e^{-h_1^{\prime }} &=&\frac{T_{ll}^{\prime }}{T_{00}^{\prime }}\,,\quad
e^{-h_2^{\prime }}=\frac{T_{ll}^{\prime }}{T_{00}^{\prime }}\left( \frac{%
T_{l0}^{\prime }}{T_{ll^{\prime }}^{\prime }}\right) ^2\,,  \label{newvl}
\end{eqnarray}
where $l,l^{\prime }=1,\ldots ,q-1$, and the results are independent of the
choice of $l\ne l^{\prime }$. From Eq. (\ref{trm}) it follows that: 
\[
\left( \hat{T}^{n+1}\right) _{00}=T_{00}\left( \hat{T}^n\right)
_{00}+(q-1)T_{l0}\left( \hat{T}^n\right) _{l0}=\left( \hat{T}^n\right)
_{00}+(q-1)ce^{-(h_1-h_2)/2}\left( \hat{T}^n\right) _{l0}, 
\]
and, analogously, 
\[
\left( \hat{T}^{n+1}\right) _{l0}=ce^{-(h_1-h_2)/2}\left( \hat{T}^n\right)
_{00}+e^{-h_1}\left[ 1+(q-2)ce^{h_2}\right] \left( \hat{T}^n\right) _{l0}\,, 
\]
\[
\left( \hat{T}^{n+1}\right) _{ll}=ce^{-(h_1-h_2)/2}\left( \hat{T}^n\right)
_{l0}+e^{-h_1}\left( \hat{T}^n\right) _{ll}+(q-2)ce^{-h_1+h_2}\left( \hat{T}%
^n\right) _{ll^{\prime }}\,, 
\]
\[
\left( \hat{T}^{n+1}\right) _{ll^{\prime }}=ce^{-(h_1-h_2)/2}\left( \hat{T}%
^n\right) _{l0}+ce^{-h_1+h_2}\left( \hat{T}^n\right) _{ll}+e^{-h_1}\left[
1+(q-3)ce^{h_2}\right] \left( \hat{T}^n\right) _{ll^{\prime }}\,. 
\]
Here the symmetry property $T_{\eta _1\eta _2}=T_{\eta _2\eta _1}$ was used.
Introducing: 
\begin{eqnarray*}
\tilde{u}(n) &=&\left. \frac \partial {\partial q}\left( \hat{T}^n\right)
_{00}\right| _{q=1}\,,\quad t_1(n)=\left. \left( \hat{T}^n\right)
_{l0}\right| _{q=1}\,, \\
t_2(n) &=&\left. \left( \hat{T}^n\right) _{ll}\right| _{q=1}\,,\quad
t_3(n)=\left. \left( \hat{T}^n\right) _{ll^{\prime }}\right| _{q=1}\,,
\end{eqnarray*}
and taking into account that at $q=1$ we have $T_{00}=\left( \hat{T}%
^n\right) _{00}=T_{00}^{\prime }=1$, the following equations can be
established to read: 
\begin{eqnarray}
t_1(n+1) &=&ce^{-(h_1-h_2)/2}+e^{-h_1}\left( 1-ce^{h_2}\right) t_1(n)\,, 
\nonumber \\
t_2(n+1) &=&ce^{-(h_1-h_2)/2}t_1(n)+e^{-h_2}t_2(n)-ce^{-h_1+h_2}t_3(n)\,, 
\nonumber \\
t_3(n+1) &=&ce^{-(h_1-h_2)/2}t_1(n)+ce^{-h_1+h_2}t_2(n)+e^{-h_1}\left(
1-2ce^{h_2}\right) t_3(n)\,,  \label{teq} \\
\tilde{u}(n+1) &=&\tilde{u}(n)+ce^{-(h_1-h_2)/2}t_1(n)\,,  \nonumber
\end{eqnarray}
with initial conditions: $t_1(0)=t_3(0)=\tilde{u}(0)=0$, $t_2(0)=1$, which
can be easily solved. After the transition to infinitesimal transformation, $%
n=1+d\lambda /\lambda $, $m=1+\epsilon d\lambda /\lambda $, we have, taking
into account Eq. (\ref{ntrm}): 
\begin{eqnarray}
T_{ol}^{\prime } &=&t_1^m(n)=ce^{-(h_1-h_2)/2}\left\{ 1+\frac{d\lambda }%
\lambda \left[ \epsilon \left( \ln c-\frac{h_1-h_2}2\right) +\right. \right.
\nonumber \\
&&\ \ \left. \left. \left( 1-\frac{e^{h_1}}{e^{h_1}+ce^{h_2}-1}\right)
\,\left[ \ln \left( 1-ce^{h_2}\right) -h_1\right] \right] \right\}, 
\nonumber \\
T_{ll}^{\prime } &=&t_2^m(n)=e^{-h_1}\left\{ 1+\frac{d\lambda }\lambda
\left[ -\epsilon h_1+ce^{h_2}\left( 1-\frac{ce^{h_1}}{e^{h_1}+ce^{h_2}-1}%
\right) +\right. \right.  \nonumber \\
&&\ \ \left. \left. \left( 1-\frac{c^2e^{2h_1+h_2}}{\left(
e^{h_1}+ce^{h_2}-1\right) ^2}\right) \,\left[ \ln \left( 1-ce^{h_2}\right)
-h_1\right] \right] \right\},  \label{izntrm} \\
T_{ll^{\prime }}^{\prime } &=&t_3^m(n)=ce^{-h_1+h_2}\left\{ 1+\frac{d\lambda 
}\lambda \left[ \epsilon \left( \ln c-h_1+h_2\right) +1-\frac{ce^{h_1}}{
e^{h_1}+ce^{h_2}-1}+\right. \right.  \nonumber \\
&&\ \ \left. \left. \left( 1-\frac{ce^{2h_1}}{\left(
e^{h_1}+ce^{h_2}-1\right) ^2}\right) \,\left[ \ln \left( 1-ce^{h_2}\right)
-h_1\right] \right] \right\},  \nonumber \\
\left. \frac \partial {\partial q}\ln T_{00}^{\prime }\right| _{q=1} &=&%
\frac{d\lambda }\lambda \frac{c^2e^{h_2}}{\left( e^{h_1}+ce^{h_2}-1\right) ^2%
}\left\{ 1-\left( 1-\frac{e^{h_1}}{e^{h_1}+ce^{h_2}-1}\right) \,\left[ \ln
\left( 1-ce^{h_2}\right) -h_1\right] \right\}.  \nonumber
\end{eqnarray}
From Eqs. (\ref{izntrm}), (\ref{izdef}) and (\ref{newvl}) it immediately
follows that: 
\begin{eqnarray}
v_c &=&\epsilon c\ln c-c\left( 1-\frac{ce^{h_1}}{e^{h_1}+ce^{h_2}-1}\right) -
\nonumber \\
&&\ \ \ c\left[ \frac{2e^{h_1}}{e^{h_1}+ce^{h_2}-1}-1-\frac{ce^{2h_1}}{%
\left( e^{h_1}+ce^{h_2}-1\right) ^2}\right] \,\left[ \ln \left(
1-ce^{h_2}\right) -h_1\right] \,,  \label{vc}
\end{eqnarray}
\begin{eqnarray}
v_1 &=&\epsilon h_1-ce^{h_1}\left( 1-\frac{ce^{h_1}}{e^{h_1}+ce^{h_2}-1}
\right) -  \nonumber \\
&&\ \ \ \left[ 1-\frac{c^2e^{2h_1+h_2}}{\left( e^{h_1}+ce^{h_2}-1\right) ^2}
\right] \left[ \ln \left( 1-ce^{h_2}\right) -h_1\right] \,,  \label{v1}
\end{eqnarray}
\begin{eqnarray}
&&v_2=\epsilon h_2+\left( 2-ce^{h_2}\right) \left( 1-\frac{ce^{h_1}}{
e^{h_1}+ce^{h_2}-1}\right) -  \nonumber \\
&&\left[ 1-\frac{c^2e^{2h_1+h_2}}{\left( e^{h_1}+ce^{h_2}-1\right) ^2}-\frac{
2e^{h_1}}{e^{h_1}+ce^{h_2}-1}\left( 1-\frac{ce^{h_1}}{e^{h_1}+ce^{h_2}-1}
\right) \right] \left[ \ln \left( 1-ce^{h_2}\right) -h_1\right] \,,
\label{v2}
\end{eqnarray}
\begin{equation}
w=\frac{c^2e^{h_2}}{\left( e^{h_1}+ce^{h_2}-1\right) ^2}\left\{ 1-\left( 1- 
\frac{ce^{h_1}}{e^{h_1}+ce^{h_2}-1}\right) \left[ \ln \left(
1-ce^{h_2}\right) -h_1\right] \right\} \,.  \label{w}
\end{equation}

\section{Evaluation of topological quantities at $\epsilon \ll 1$.}

\label{apb}

At $c\ll 1$ Eq.(\ref{bbev}) may be rewritten as: 
\[
\lambda \frac{dc}{d\lambda }=\epsilon c\ln \frac c{c_t}\,, 
\]
which can be easily solved to yield: 
\begin{equation}
c=c_t\exp \left[ \pm \left( \frac \lambda {\lambda _0}\right) ^\epsilon
\right] \,,  \label{solc1}
\end{equation}
where $\lambda _0$ is the arbitrary positive integration constant. On the
other hand, one can neglect first term in Eq.(\ref{bbev}) when $c\gg c_t$,
obtaining the solution: 
\begin{equation}
c=1-\exp (-\lambda /\lambda _1)\,,  \label{solc2}
\end{equation}
with the other integration constant $\lambda _1$.

One can match plus sign solution (\ref{solc1}) with Eq. (\ref{solc2}) in the
region $c_t\ll c\ll 1$. Setting $\lambda _0=\epsilon ^{1/\epsilon }\tilde{%
\lambda}_0$ in Eq. (\ref{solc1}), one can see, taking into account Eq. (\ref
{thrn1d}), that to fulfill $1\gg c\gg c_t$, one should require $0<1-(\lambda
/\tilde{\lambda}_0)^\epsilon \ll 1$; therefore one can set ($\lambda /\tilde{%
\lambda}_0)^\epsilon \approx 1+\epsilon \ln (\lambda /\tilde{\lambda}_0)$,
and formula (\ref{solc1}) turns into $c\approx \lambda /\tilde{\lambda}_0$.
On the other hand, we have from Eq. (\ref{solc2}), that within the same
region $c\approx \lambda /\lambda _1$, that is, $\tilde{\lambda}_0\approx
\lambda _1$, or $\lambda _0\approx \epsilon ^{1/\epsilon }\lambda _1$.

When $c\ll 1$, taking into account (\ref{solc1}), Eq. (\ref{efe0}) reads (we
set the integration constant $\lambda _0=1$): 
\begin{equation}
\lambda \frac{df}{d\lambda }-(1+\epsilon )f=\frac{c_t^2}2\exp \left( \pm
2\lambda ^\epsilon \right) \,.  \label{efe0a}
\end{equation}
At $c<c_t$ (minus sign in the right hand side of Eq. (\ref{efe0a})), one
should set $f=0$ at $\lambda \rightarrow \infty $, and the corresponding
solution is: 
\begin{eqnarray}
f(\lambda ) &=&\frac{c_t^2}{2\epsilon }\left( 2\lambda ^\epsilon \right) ^{%
\frac 1\epsilon +1}\Gamma \left( -\frac 1\epsilon -1,2\lambda ^\epsilon
\right)  \nonumber \\
&=&-\frac 12\frac{c_t^2}{1+\epsilon }\frac{\pi \epsilon }{\sin \frac \pi %
\epsilon }\frac{\left( 2\lambda ^\epsilon \right) ^{\frac 1\epsilon +1}}{
\Gamma \left( \frac 1\epsilon \right) }-\frac{c_t^2}2\sum_{n=0}^\infty \frac{%
\left( -2\lambda ^\epsilon \right) ^n}{n![1-\epsilon (n-1)]},  \label{febel}
\end{eqnarray}
where $\Gamma (a,x)$ is the incomplete $\Gamma $-function \cite{be}, tending
to 0 at $x\rightarrow \infty $. Substituting in the first term in the right
hand side of formula (\ref{febel}) $\lambda ^\epsilon =\ln (c_t/c)$, and
replacing $\Gamma (1/\epsilon )$ with its asymptotic expression, one arrives
at Eq. (\ref{fe0s}). At $1\gg c>c_t$, the general solution of Eq. (\ref
{efe0a}) with a plus sign is: 
\begin{equation}
f(\lambda )=-\frac{c_t^2}{2\epsilon }\left( -2\lambda ^\epsilon \right) ^{%
\frac 1\epsilon +1}\gamma \left( -\frac 1\epsilon -1,-2\lambda ^\epsilon
\right) -A\lambda ^{1+\epsilon }\,,  \label{efe0b}
\end{equation}
where $\gamma (a,x)=\Gamma (a)-\Gamma (a,x)$ (note that $x^{-a}\gamma (a,x)$
is analytic function of $x$ at $x=0$), and $A$ is an integration constant to
be determined through combining with the expression for the ``free energy''
at $c\gg c_t$. The latter may be done, substituting Eq. (\ref{solc2}) with $%
\lambda _1=\epsilon ^{1/\epsilon }$ into right hand side of Eq.(\ref{efe0}).

Having in mind that solutions (\ref{febel},\ref{efe0}) are to be matched
with each other in lowest order on $\epsilon $, and setting $\epsilon =0$ in
the left-hand side of Eq.(\ref{efe0}) too, we arrive at: 
\begin{equation}
f=\exp \left( -\epsilon ^{1/\epsilon }\lambda \right) -1=-c\,.  \label{efe0c}
\end{equation}
On the other hand, from Eq.(\ref{efe0b}) we have within the same region,
replacing the function $\gamma $ with its asymptotics at a large value of
its second argument \cite{be}: 
\begin{equation}
f\approx -\frac{c_t^2}{4\epsilon \lambda ^\epsilon }\exp \left( 2\lambda
^\epsilon \right) -A\lambda ^{1+\epsilon }\approx -\frac{c^2}{4\epsilon \ln 
\frac c{c_t}}-A\epsilon ^{-1/\epsilon -1}c\,.  \label{efe0d}
\end{equation}
Comparing expressions (\ref{efe0c}) and \ref{efe0d}), one can conclude that $%
A\approx \epsilon ^{1/\epsilon +1}$. Substituting this into (\ref{efe0b}),
replacing there $\lambda ^\epsilon $ with $\ln (c/c_t)$, and combining with
Eqs. (\ref{febel}) and (\ref{efe0c}), one arrives at Eq. (\ref{fe0}).

\section{Inverse Laplace transformation of the distribution function.}

\label{apc}

After some integrations by parts, Eq.(\ref{leftx1d}) may be rewritten as: 
\begin{eqnarray}
\ln x\left( \phi \right) &=&i\pi +\left( 1-\frac a{1+\epsilon }\right) \ln
\left( -\phi \right) +\frac a{1+\epsilon }\left( 1-\gamma -\ln \left(
1+\epsilon \right) +\frac 1{\left( 1+\epsilon \right) \phi }\right) + 
\nonumber \\
&&\ \ \ \ \frac{e^\phi -1-\phi }\phi +\frac a{\left( 1+\epsilon \right) ^2}%
\int_{-\infty }^\phi \frac{d\zeta }{\zeta ^2}e^{\left( 1+\epsilon \right)
\zeta }  \nonumber \\
\ &\approx &i\pi +\left( 1-\frac a{1+\epsilon }\right) \ln \left( -\phi
-\phi _1\right) +\frac a{1+\epsilon }G_1+c_t\frac{e^\phi }\phi +\frac a{%
\left( 1+\epsilon \right) ^3}\frac{e^{\left( 1+\epsilon \right) \phi }}\phi,
\label{aslnx}
\end{eqnarray}
$\phi _1=a/\left( 1+\epsilon \right) ^2$, $G_1=1-\gamma -\ln \left(
1+\epsilon \right) -\left( 1+\epsilon \right) c_t/a$. From the letter
expression in Eq.(\ref{aslnx}), which is the asymptotics as $\left| \phi
\right| \gg 1$, we have: 
\begin{equation}
x\left( \phi \right) =-e^{\frac a{1+\epsilon }G_1}\left( -\phi -\phi
_1\right) ^{1-\frac a{1+\epsilon }}+c_te^\phi +\frac a{\left( 1+\epsilon
\right) ^3}\frac{e^{\left( 1+\epsilon \right) \phi }}\phi \,.  \label{asx}
\end{equation}
In the evaluation of the inverse Laplace transformation (\ref{scf}) one can
neglect two last terms in Eq.(\ref{asx}) if one considers the DF $\bar{\Pi}%
\left( y\right) $ at $y$ close enough to $1$. Substituting $x\left( \phi
\right) $ into Eq.(\ref{scf}), rescaling integration variable as: 
\[
-\phi -\phi _1=e^{G_1}y^{\frac{1+\epsilon }a}u\equiv \frac{1+\epsilon }a
Wu\,, 
\]
and taking into account $a\ll 1$, which leads to the following expression in
the exponent in Eq.(\ref{scf}): 
\[
-\phi +yx\left( \phi \right) =\phi _1+\frac{1+\epsilon }aWu\left( 1-u^{-%
\frac a{1+\epsilon }}\right) \approx \phi _1+Wu\ln u\,, 
\]
one immediately arrives at the formulas (\ref{a1daslarge}--\ref{sfunc}).

To evaluate $S\left( W\right) $, it is convenient to use the integration
contour $\Im \left( u\ln u\right) =0$, or, introducing polar coordinates $%
u=r\exp \left( i\theta \right) $: 
\begin{equation}
r=\exp \left( -\theta \cot \theta \right) \,.  \label{icfc}
\end{equation}
Substituting Eq.(\ref{icfc}) into Eq.(\ref{sfunc}), we have: 
\begin{equation}
S\left( W\right) =\int_{-\pi }^\pi \frac{d\theta }{2\pi }V(\theta
)e^{-WV(\theta )}\,,\;\;V(\theta )=\frac \theta {\sin \theta }e^{-\theta
\cot \theta }\,.  \label{sfunc1}
\end{equation}
The value of the integral may be estimated by the saddle-point method,
looking for the maxima of the expression $\ln V\left( \theta \right)
-WV\left( \theta \right) $. The stationary point equation is: 
\begin{equation}
V^{\prime }\left( \theta \right) \left[ W-\frac 1{V\left( \theta \right) }%
\right] =0\,.  \label{stpt}
\end{equation}

As $W\gg 1$, the stationary point is $\theta _s=0$, and the asymptote of Eq.(%
\ref{sfunc1}) is: 
\begin{equation}
S\left( W\right) \approx \left( 2\pi eW\right) ^{-1/2}\exp \left(
-W/e\right) .  \label{assc1}
\end{equation}
On the other hand, if $W\ll 1$, we have two stationary points $\pm \theta _s$%
, $V\left( \pm \theta _s\right) =1/W$, $\theta _s=\pi -\delta $, and for $%
\delta \ll 1$ the following equation may be obtained: 
\[
\frac \pi \delta \exp \left( \frac \pi \delta -1\right) =\frac 1W\,, 
\]
the solution for which is: 
\[
\pi /\delta \approx \frac{\ln \left( e/W\right) }{\ln \ln \left( e/W\right) }%
\,. 
\]
The asymptotics of $S\left( W\right) $ turns out to be: 
\begin{equation}
S\left( W\right) \approx \frac{\sqrt{2\pi }}{eW}\left[ \frac{\ln \ln \left(
e/W\right) }{\ln \left( e/W\right) }\right] ^2\,.  \label{assc2}
\end{equation}

\newpage

\begin{table}[tbp] \centering%
\begin{tabular}{ccccccc}
D &  & $1+\epsilon $ & 2 & 3 & 4 & $D\rightarrow \infty $ \\ 
&  &  &  &  &  &  \\ 
$c_t$ & 
\begin{tabular}[t]{c}
RGMK \\ 
Other
\end{tabular}
& 
\begin{tabular}[t]{c}
$e^{-1/\epsilon }$ \\ 
---
\end{tabular}
& 
\begin{tabular}[t]{c}
1/2 \\ 
1/2
\end{tabular}
& 
\begin{tabular}[t]{l}
0.840 \\ 
\begin{tabular}{c}
0.751(BP) \\ 
0.83(CP)
\end{tabular}
\end{tabular}
& 
\begin{tabular}[t]{c}
0.945 \\ 
0.840(BP)
\end{tabular}
& 
\begin{tabular}[t]{c}
1 \\ 
1
\end{tabular}
\\ 
&  &  &  &  &  &  \\ 
$\nu $ & 
\begin{tabular}{c}
RGMK \\ 
Other
\end{tabular}
& 
\begin{tabular}{c}
$1/\epsilon $ \\ 
---
\end{tabular}
& 
\begin{tabular}{c}
1.629 \\ 
4/3=1.333
\end{tabular}
& 
\begin{tabular}{c}
1.219 \\ 
0.89
\end{tabular}
& 
\begin{tabular}{c}
1.092 \\ 
0.68
\end{tabular}
& 
\begin{tabular}{c}
1 \\ 
1/2
\end{tabular}
\\ 
&  &  &  &  &  &  \\ 
$\beta $ & 
\begin{tabular}{c}
RGMK \\ 
Other
\end{tabular}
& 
\begin{tabular}{c}
$\frac 1{3\epsilon }e^{-2/\epsilon }$ \\ 
---
\end{tabular}
& 
\begin{tabular}{c}
0.137 \\ 
5/36=0.139
\end{tabular}
& 
\begin{tabular}{c}
0.255 \\ 
0.40
\end{tabular}
& 
\begin{tabular}{c}
0.325 \\ 
0.65
\end{tabular}
& 
\begin{tabular}{c}
$\frac 1D$ \\ 
1
\end{tabular}
\\ 
&  &  &  &  &  &  \\ 
$\mu $ & 
\begin{tabular}{c}
RGMK \\ 
Other
\end{tabular}
& 
\begin{tabular}{c}
$\frac{1+\epsilon }{\epsilon ^2}\exp \left( -\frac{1+\epsilon }\epsilon
\right) $ \\ 
---
\end{tabular}
& 
\begin{tabular}{c}
1.333 \\ 
1.303
\end{tabular}
& 
\begin{tabular}{c}
2.330 \\ 
2.00
\end{tabular}
& 
\begin{tabular}{c}
3.230 \\ 
2.39
\end{tabular}
& 
\begin{tabular}{c}
$D-1$ \\ 
3
\end{tabular}
\end{tabular}
\caption{Threshold values $c_t$ for bond (BP) and continuous (CP) percolation
and critical exponents evaluated by the RGMK
method of this work compared to the exact (if known) or best possible numeric values.
Values for comparison were extracted from reviews \protect\cite{nyo94,imb92},
for the conductivity exponent $\mu$ after Ref. \protect\cite{nh97}.
} \label{exponents}%
\end{table}%

\end{document}